\documentclass[%
preprint,
11pt,
 amsmath,amssymb,
aps,
]{revtex4-2}

\usepackage{graphicx}
\usepackage{epstopdf}
\usepackage{dcolumn}
\usepackage{bm}
\usepackage{xcolor}
\usepackage{upgreek}
\usepackage{multirow}
\usepackage{tabularx}
\usepackage[abs]{overpic}
\usepackage{placeins}
\usepackage{tikz}
\usepackage{subfig}
\usepackage{comment}

\begin{document}


\title{When is settling important for particle concentrations in wall-bounded turbulent flows?}

\author{A.D. Bragg}
\email{andrew.bragg@duke.edu}
\affiliation{Department of Civil and Environmental Engineering, Duke University, Durham, NC 27708, USA}

\author{D. H. Richter}%
\affiliation{%
Department of Civil and Environmental Engineering and Earth Sciences, University of Notre Dame, Notre Dame, IN 46556, USA
}%

\author{G. Wang}%
\email{gwang4academy@gmail.com}
\affiliation{%
Physics of Fluids Group and Twente Max Planck Center, Department of Science and Technology, Mesa+ Institute, and J. M. Burgers Center for Fluid Dynamics, University of Twente, P.O. Box 217, 7500 AE Enschede, The Netherlands
}%

\date{\today}

\begin{abstract}

We explore the role of gravitational settling on inertial particle concentrations in a wall-bounded turbulent flow. While it may be thought that settling can be ignored when the settling parameter $Sv\equiv v_s/u_\tau$ is small ($v_s$ - Stokes settling velocity, $u_\tau$ - fluid friction velocity), we show that even in this regime the settling may make a leading order contribution to the concentration profiles. This is because the importance of settling is determined, not by the size of $v_s$ compared with $u_\tau$ or any other fluid velocity scale, but by the size of $v_s$ relative to the other mechanisms that control the vertical particle velocity and concentration profile. We explain this in the context of the particle mean-momentum equation, and show that in general, there always exists a region in the boundary layer where settling cannot be neglected, no matter how small $Sv$ is (provided it is finite). Direct numerical simulations confirm the arguments, and show that the near-wall concentration is highly dependent on $Sv$ even when $Sv\ll 1$, and can reduce by an order of magnitude when $Sv$ is increased from $O(10^{-4})$ and $O(10^{-2})$. The results also show that the preferential sampling of ejection events in the boundary layer by inertial particles when $Sv=0$ is profoundly altered as $Sv$ is increased, and is replaced by a preferential sampling of sweep events due to the onset of the preferential sweeping mechanism.

\end{abstract}

\maketitle

\section{Introduction}

In many particle-laden, wall-bounded turbulent flows, it is important to characterize the near-wall distribution of a dispersed phase. In the environment, surface emission of particulate matter (e.g. dust, aerosols) is often estimated by assuming a relationship between the mean concentration and surface flux. This flux-profile relationship is also the basis for wall models of heavy scalar transport \cite{ChameckiAS2009}.

While many past studies have focused on the phenomena of turbophoresis \cite{reeks83,johnson20} and the interactions of particles with near-wall coherent structures \cite{rashidi90,eaton94,SoldatiIJMF2009}, much less attention has been given to the settling of particles through wall-bounded turbulence and the impact this has on the particle concentrations. Indeed, although it is well-known that particles can experience an inertial enhancement to their settling velocity in isotropic, homogeneous turbulence \cite{maxey87, wang1993settling}, the interplay between settling and inertial effects, especially near the wall, is much less understood. In a recent study we explored this question using theory and direct numerical simulations (DNS) to consider how the mechanisms governing the particle settling and concentration profiles vary with distance from the wall \cite{bragg2021mechanisms}. It was shown that sufficiently far from the wall, the main mechanism that modifies the settling compared to the Stokes settling velocity is the enhancement due to the preferential sweeping mechanism of \citet{maxey87}. Closer to the wall, however, this mechanism becomes subleading and instead the turbophoretic drift velocity \cite{reeks83} dominates the drift of the particles towards the wall. These changes in the mechanisms controlling the particle settling velocity directly impact the particle concentration profiles, since the two quantities are connected via the particle continuity equation \cite{bragg2021mechanisms}. These results highlight the particular mechanisms that models which account for settling, but ignore particle inertia (e.g. \citet{rouse37,li2021cospectral}) must incorporate if they are to be extended to the regime of finite particle inertia.

In many studies that are aimed at understanding the behavior of inertial particles in wall-bounded turbulence, the effect of gravity is often neglected, usually under the assumption that the settling velocity is small. The relevant nondimensional parameter which describes this, however, is often assumed to be the ratio of the Stokes settling velocity $v_s\equiv\tau_{p} g$ (where $\tau_{p}$ is the particle response time and $g$ is the gravitational acceleration) to the flow friction velocity $u_{\tau}$. What we aim to demonstrate both theoretically and numerically, however, is that near the wall, even small Stokes settling velocities can be comparable or larger than the other contributions to the particle settling velocity (e.g the turbophoretic velocity) that generate concentration buildup near the wall. This in turn disrupts commonly assumed balances which ultimately determine the wall-normal distribution of the particles.

\section{Theoretical analysis}\label{TandA}
We consider the vertical motion of small, heavy particles subject to Stokes drag and gravitational forces \cite{maxey83}
\begin{align}
\frac{d}{dt}w^p(t)=\frac{1}{\tau_p}\Big(u^p(t)-w^p(t)\Big)-g,
\end{align}
where $w^p(t)$ is the vertical particle velocity and $u^p(t)$ is the vertical fluid velocity at the particle position. The system will also be assumed to have sufficiently low volume fraction so that one-way coupling can be assumed together with the absence of particle-particle collisions. 

In assessing the importance of gravitational settling on particle motion in turbulent flows, it is typical to define a settling parameter such as $\tau_p g/u_\eta$ \cite[where $u_\eta$ is the Kolmogorov velocity scale]{BALACHANDAR09} or in the context of boundary layers $\tau_p g/u_\tau$ \cite{johnson20}, and then to conclude that the effect of settling can be neglected when these non-dimensional parameters are small. Nevertheless, in some works, the effect of gravitational settling has been shown to be important even when such parameters are small. For example, DNS results in \citet{richter18,bragg2021mechanisms} showed that even when $\tau_p g/u_\tau=O(10^{-2})$, the effect of gravitational settling on the particle motion in the boundary layer was strong, and in some parameter regimes, made a leading order contribution to the particle motion. As we will now show, this is because quantities such as $\tau_p g/u_\tau$ are inappropriate measures of the importance of settling on the particle motion in the boundary layer.
  
In the following, all quantities are normalized using the fluid friction timescale $\tau$ and velocity scale $u_\tau$ to express them in wall units, usually denoted by the superscript $+$. However, in what follows, we drop the superscript for notational simplicity.   
  
Using phase-space Probability Density Function (PDF) equations, we can construct transport equations for the average concentration $\varrho\equiv \langle \delta (z^p(t)-z)\rangle$, where $z^p(t)$ is the vertical particle position, $z$ is the time-independent vertical position coordinate (with $z=0$ corresponding to the wall), and $\langle\cdot\rangle_z$ denotes an ensemble average conditioned on $z^p(t)=z$. The continuity equation governing $\varrho(z,t)$ is \cite{bragg2021mechanisms}
\begin{align}
\partial_t\varrho+\nabla_z[\varrho \langle w^p(t)\rangle_z]=0.\label{CE}    
\end{align}
Equation \eqref{CE} can be solved by specifying $ \langle w^p(t)\rangle_z$, for which an expression can be obtained from the particle mean-momentum equation \cite{bragg2021mechanisms}
\begin{align}
 \langle w^p(t)\rangle_z=\langle{u}^p(t)\rangle_{{z}}-Sv-St( D_t\langle w^p(t)\rangle_z + \nabla_z \mathcal{W}+\mathcal{W}\varrho^{-1}\nabla_z\varrho),\label{weq}
\end{align}
where $D_t\equiv \partial_t+\langle w^p(t)\rangle_z\nabla_z $, $\mathcal{W}\equiv \langle (w^p(t)-\langle{w}^p(t)\rangle_{{z}})^2\rangle_z$ is the variance of the vertical particle velocity, $St\equiv \tau_p/\tau$ is the Stokes number, $Sv\equiv \tau_p g/u_\tau$ is the settling number. Hereafter, the mean vertical velocity $\langle w^p(t)\rangle_z$ will sometimes be referred to as the settling velocity, which is to be distinguished from the Stokes settling velocity $Sv$.  

The term $\langle{u}^p(t)\rangle_{{z}}$ in \eqref{weq} is a mean velocity that arises from the particles preferentially sampling the underlying turbulent flow \cite{johnson20}, that vanishes for fully-mixed fluid particles \cite{bragg12b}. The term $D_t\langle w^p(t)\rangle_z$ is the mean particle acceleration. The term $-St\nabla_z \mathcal{W}$ is the turbophoretic velocity which arises due to the combination of turbulence inhomogeneity and particle inertia \cite{reeks83}. Finally, the term $-St\mathcal{W}\varrho^{-1}\nabla_z\varrho$ is a diffusive velocity that arises from decoupling between the particle velocity and the local fluid velocity, and is only finite when the concentration field is non-uniform. Detailed explanations of each of the terms in \eqref{weq} may be found in \citet{bragg2021mechanisms}.


We will focus on the steady-state regime for which the solution to \eqref{CE} is that the flux $\varrho \langle w^p(t)\rangle_z$ is a constant (that depends upon the boundary conditions and system parameters). For $Sv=0$, a possible steady-state is the zero-flux configuration, for which $\langle w^p(t)\rangle_z=0$. For $Sv>0$, unless resuspension mechanisms at the wall are sufficiently strong to overcome the weight of the particles, a zero-flux configuration will not be established, and instead a constant, negative flux will be established with $\langle w^p(t)\rangle_z<0$. In this paper we consider this constant negative flux regime, although the analysis and its implications could easily be extended to the zero-flux configuration.

The regime $St\ll 1$ but finite $Sv$ corresponds to the regime of negligible particle inertia, but non-negligible settling that was analyzed by \citet{rouse37}. Our interest, by contrast, is the regime $Sv\ll1$ and finite $St$, and to understand whether in this regime the contribution from $Sv$ to the total vertical velocity $\langle w^p(t)\rangle_z$ in \eqref{weq} may be ignored. When the contribution of $Sv$ to $\langle w^p(t)\rangle_z$ can be ignored, then it follows from \eqref{CE} that the concentration profile will be independent of $Sv$. Equation \eqref{weq} is regularly perturbed with respect to $Sv$, i.e. the limit of \eqref{weq} for $Sv\to 0$ is equal to \eqref{weq} when setting $Sv= 0$. To explore if $Sv$ may be neglected if it is small but non-zero we introduce
\begin{align}
 \Lambda(z)&\equiv \langle{u}^p(t)\rangle_{{z}}-St( D_t\langle w^p(t)\rangle_z + \nabla_z \mathcal{W}+\mathcal{W}\varrho^{-1}\nabla_z\varrho),\label{Lambda_eq}
\end{align}
so that $\langle w^p(t)\rangle_z=\Lambda-Sv$, and expand $\Lambda$ in $Sv$ to obtain
\begin{align}
\Lambda(z)&=\sum_{n=0}^{\infty}Sv^n\Lambda_n(z),\\
 \langle w^p(t)\rangle_z&=\Lambda_0(z)+Sv(\Lambda_1(z)-1)+O(Sv^2),    
\end{align}
where $\Lambda_0(z)=\Lambda(z)\vert_{Sv=0}$. For settling to be ignored we require that $\Lambda_0(z)\gg O(Sv)$. To consider when this condition is satisfied, we will first focus on the regime $z\ll1$ where analytical results are possible, and which is also the regime of most interest since this is where the particle concentration is highest.  It is also worth emphasizing that although the following analysis is strictly for the limited regime $z\ll1$, the asymptotic results to be used in the analysis are known to hold up to $z=O(10)$ \cite{sikovsky14,johnson20}, and therefore our results should also apply up to $z=O(10)$.

To show whether the condition $\Lambda_0(z)\gg O(Sv)$ is satisfied, we must describe each of the terms contributing to $\Lambda_0(z)$ in the near wall region. It was shown in \citet{sikovsky14} using asymptotic analysis that for $z\ll 1$ and $Sv=0$, $\varrho \sim z^{-\gamma}$ (in what follows we ignore the coefficients in the asymptotic relationships since it is the dependence on $z$ that will be of interest) where $\gamma=\max[\alpha_0,3]$, and $\alpha_0(St)\in[0,4]$ so that $\gamma(St)\in[3,4]$. The asymptotic analysis also shows that for sufficiently large $St$, an additional contribution in the expansion for $\varrho$ becomes important which generates $\varrho\sim$ constant in the limit $St\to \infty$, corresponding to particles moving ballistically through the flow. For simplicity, we will restrict our focus in this analysis to low to moderate inertia particles for which  $\varrho \sim z^{-\gamma}$ describes the correct behavior. Since $\varrho \langle w^p(t)\rangle_z$ is constant, the result $\varrho \sim z^{-\gamma}$ then implies $\langle w^p(t)\rangle_z\sim  z^{\gamma}$ and $D_t\langle w^p(t)\rangle_z\sim  z^{2\gamma -1}$. Using these results, we will now consider whether the condition $\Lambda_0(z)\gg O(Sv)$ is satisfied for weak to moderate inertia particles.

In the weak inertia regime $St\ll 1$ with $Sv=0$, $\mathcal{W}\sim z^4$ \cite{johnson20}. Furthermore, the model in \citet{sikovsky14} for $\langle{u}^p(t)\rangle_{{z}}$ yields the behavior $\langle{u}^p(t)\rangle_{{z}}\sim z^{3}$ for arbitrary $St$ in the regime where $\varrho \sim z^{-\gamma}$. Therefore, substituting these asymptotic results into the definition of $\Lambda_0$ we obtain

\begin{align}
\Lambda_0(z)\sim z^3 - St(z^{2\gamma-1}+z^3 +z^3).\label{PsiStll1}
\end{align}
For $St\ll1$, $\gamma= 3$  \cite{sikovsky14} and so \eqref{PsiStll1} has the limiting form $\Lambda_0\sim z^{3}$. This shows that the condition $\Lambda_0(z)\gg O(Sv)$, which must be satisfied if settling is to be ignored, will be violated for $z\to 0$ if $Sv>0$. Hence for $St\ll1$ there will always exist a region near the wall where gravitational settling cannot be ignored, even if $Sv\ll 1$ (but finite). 


For particles with moderate inertia and for $z\ll 1$, \citet{sikovsky14} showed that the particle velocity moments obey the asymptotic result $\langle(w^p(t)-\langle w^p(t)\rangle_z)^n\rangle_z\sim  z^{\gamma}$ for $n\geq 2$, and therefore $\mathcal{W}\sim z^\gamma$. Using this we obtain
\begin{align}
\Lambda_0(z)\sim z^3 - St(z^{2\gamma-1}+z^{\gamma-1}+z^{\gamma-1}).
\end{align}
Since $\gamma\in[3,4]$, the leading behavior is $\Lambda_0\sim z^{\gamma-1}$. Therefore, just as for the $St\ll1$ case, this also shows that the condition $\Lambda_0(z)\gg O(Sv)$ will be violated for moderately inertial particles for $z\to 0$ if $Sv>0$. 

We have therefore shown that for both weakly and moderately inertial particles, even if $Sv\ll1$, there will always exist a region close to the wall where settling cannot be ignored, and indeed where settling makes a leading order contribution to $\langle w^p(t)\rangle_z$ and therefore $\varrho$. Further way from the wall it is also possible that the condition $\Lambda_0(z)\gg O(Sv)$ could be violated since $\Lambda_0(z)$ depends on gradients in the flow statistics, and these become weak for sufficiently large $z$ (at least if the flow Reynolds number is large enough for a quasi-homogeneous region to emerge sufficiently far from the wall). To explore the role of settling on the particle dynamics throughout the boundary layer, we will now consider DNS results for inertial particle motion in an open channel flow. These results will enable us to go beyond the analysis in this section by allowing for a quantitative assessment of the impact of settling on the particle concentration profiles in the regime $Sv\ll1$.

\FloatBarrier
\section{Direct Numerical Simulations}\label{sec:DNS}

\begin{figure}
\centering
{\hspace{-5mm}\begin{overpic}
[width=0.6\textwidth,trim={100mm 0mm 0mm 8.5mm}, clip]{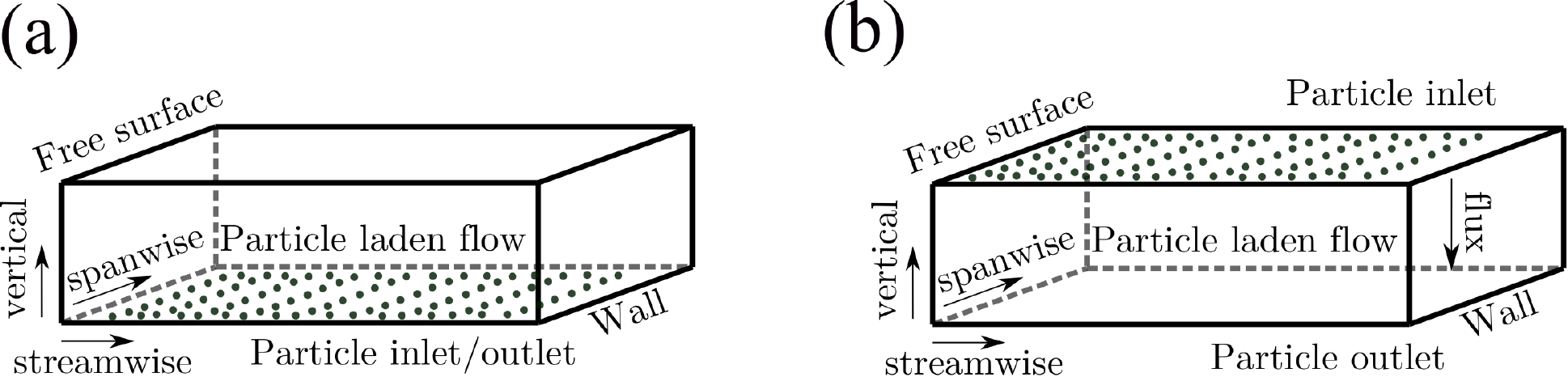}
\end{overpic}
}
\caption{Particles are initialized at the top of the domain and settle down. When they reach the wall they are removed, and replaced by another particle that is introduced at the top of the domain.}
\label{fig:configuration}
\end{figure}

DNS is used to solve the incompressible Navier-Stokes equations, which are then used in a one-way coupled scenario to solve for the motion of heavy, inertial, point-particles that are subject to drag and gravitational forces via the equation of motion
\begin{align}
\frac{d^2}{dt^2}\boldsymbol{x}^p(t)\equiv\frac{d}{dt}\boldsymbol{v}^p(t)=\frac{1}{\tau_p}\Big(\boldsymbol{u}^p(t)-\boldsymbol{v}^p(t)\Big)-\boldsymbol{g}, \label{eom}
\end{align}
where $\boldsymbol{x}^p(t),\boldsymbol{v}^p(t)$ are the particle position and velocity, $\boldsymbol{u}^p(t)$ is the fluid velocity at the particle position, and $\boldsymbol{g}$ is the gravitational acceleration in the vertical direction. The flow has a friction Reynolds number of $Re_{\tau} = 315$, and is generated by applying a constant streamwise pressure gradient to force the flow. The streamwise $x$ and spanwise $y$ directions are periodic, and the wall at $z = 0$ imposes a no-slip condition on the fluid velocity field. At the upper wall, $z = 1$, a free-slip (i.e., zero-stress) condition is imposed on the fluid velocity. The domain size is $L_x \times L_y \times L_z=2\pi \times 2\pi \times 1$ with a corresponding grid of $N_x \times N_y \times N_z=128\times256\times128$. The grid is stretched in the wall-normal direction and thus the simulations have a resolution of $\Delta^+_x \times \Delta^+_y \times \Delta^+_z=15.4 \times 7.73 \times 0.25(wall), 4.49(center)$. The unladen flow field from DNS has been tested and validated by comparison with published data for $Re_\tau=40-950$ \cite{wang2019modulation, wang_richter_2019}. Inertial particles are introduced into the flow at $z=1$ with an initial velocity equal to the Stokes settling velocity plus the local fluid velocity. When they eventually settle to the wall at $z=0$ they are removed and replaced by another particle injected at $z=1$. This system is then simulated until the flow reaches steady-state. The point-particle DNS code has been validated for inertial particles in the range $St=30-2000$ \cite{wang2019inertial}. This setup provides a canonical case of steady-state, particle-laden wall-bounded turbulence which provides a uniform downwards flux of particle concentration; figure \ref{fig:configuration} provides a schematic of the configuration.


\FloatBarrier


In order to test the idea presented in the previous section, it would be desirable to compare results for $Sv=0$ with those from $0<Sv\ll 1$ in order to see the impact of settling when $Sv$ is small. However, at finite $St$ and with the absorbing wall boundary condition described above, particles accumulate in the vicinity of the wall and free-surface for simulations using $Sv=0$, since there exists no gravitational settling with which to exit the laminar viscous layer, and resuspension events are rare. This results in a situation where the simulations cannot reach a steady state in a reasonable amount of time. Therefore, we instead consider results for finite $Sv$ in the range $0<Sv\ll 1$. If the results are insensitive to $Sv$ then this would show that settling can be ignored in this regime. If, however, there are regions of the flow where the results are sensitive to $Sv$ even when $Sv\ll 1$, then this would confirm the prediction of the theoretical analysis in the previous section. In view of this, we consider simulations with $Sv= 3 \times 10^{-4}, 3 \times 10^{-3}, 3 \times 10^{-2}, 3 \times 10^{-1}$, and for each of these we consider four different Stokes numbers $St= 0.93, 2.80, 9.32, 46.67$. Roughly speaking, this value of $Sv$ would correspond to a sand/dust grain with a diameter of $O(1-100~\mathrm{\mu m})$ suspended over a windy surface (e.g. $u_{\tau} \approx O(0.1~\mathrm{m/s})$). Note that 
$Sv= 3 \times 10^{-1}$ could be considered to be too large to satisfy the requirement $Sv\ll1$; we nevertheless include it to better identify the limits of the above theory.

\section{Results \& Discussion}
\FloatBarrier
\begin{figure}
\centering
\subfloat[]{}
\begin{overpic}
[trim = 0mm 60mm 0mm 70mm,scale=0.3,clip,tics=20]{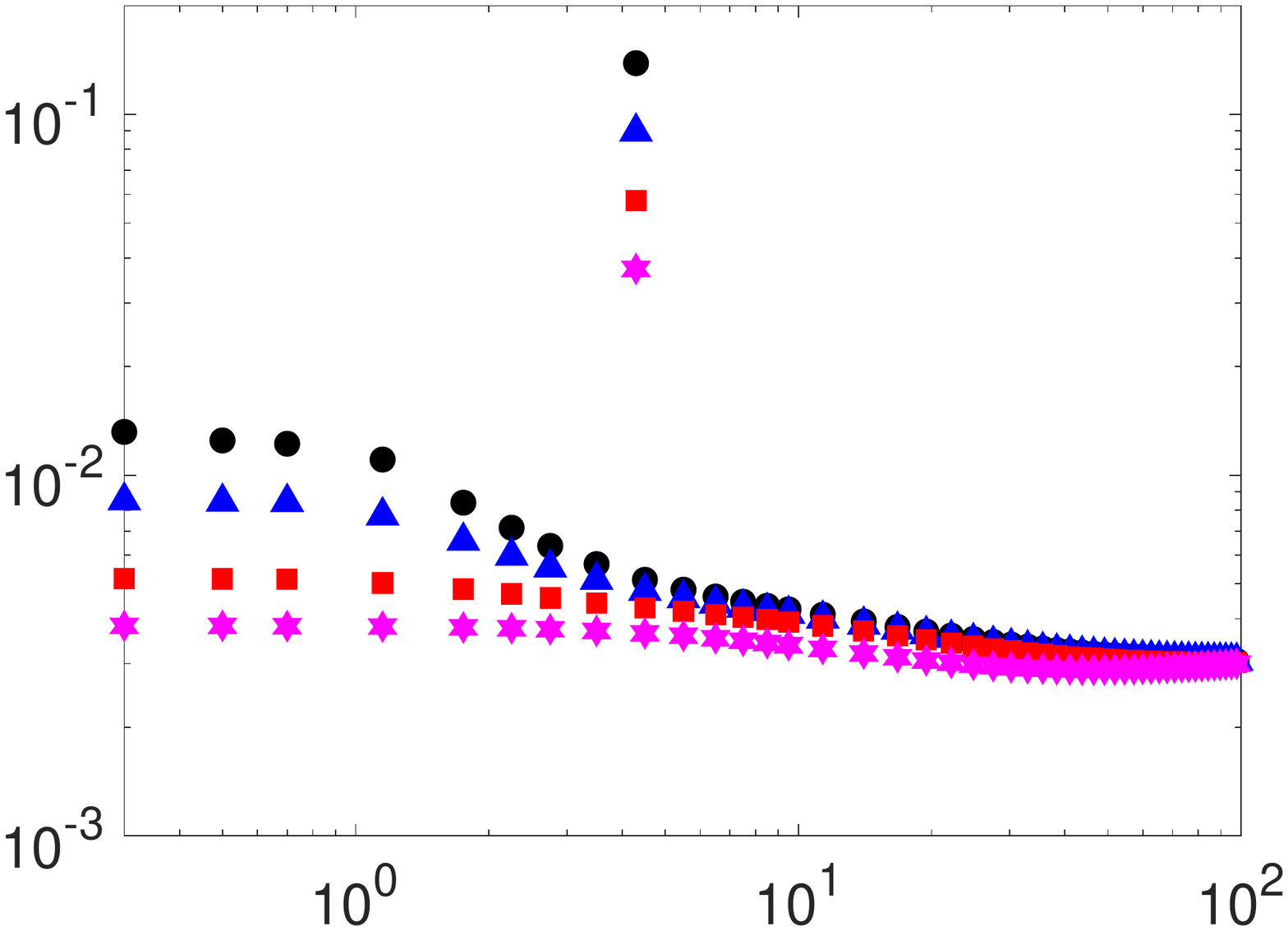}
\put(90,0){$z$}
\put(0,60){\rotatebox{90}{$\varrho$}}
\put(93,114){{\small$Sv=3\times 10^{-4}$}}
\put(93,105){{\small$Sv=3\times 10^{-3}$}}
\put(93,96){{\small$Sv=3\times 10^{-2}$}}
\put(93,87){{\small$Sv=3\times 10^{-1}$}}
\end{overpic}
\subfloat[]{}
\begin{overpic}
[trim = 0mm 60mm 0mm 70mm,scale=0.3,clip,tics=20]{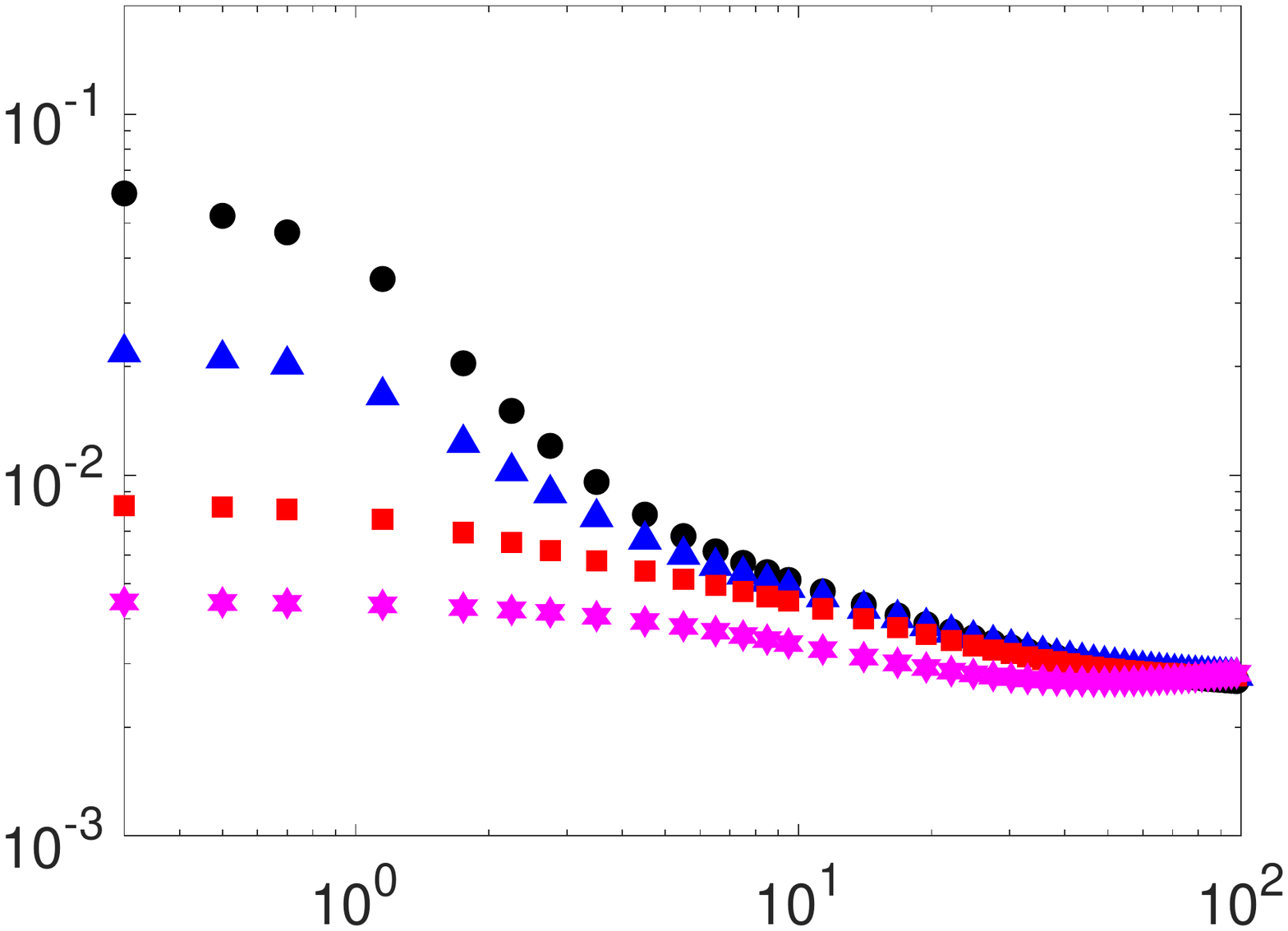}
\put(90,0){$z$}
\put(0,60){\rotatebox{90}{$\varrho$}}
\end{overpic}\\\vspace{-4mm}
\subfloat[]{}
\begin{overpic}
[trim = 0mm 60mm 0mm 50mm,scale=0.3,clip,tics=20]{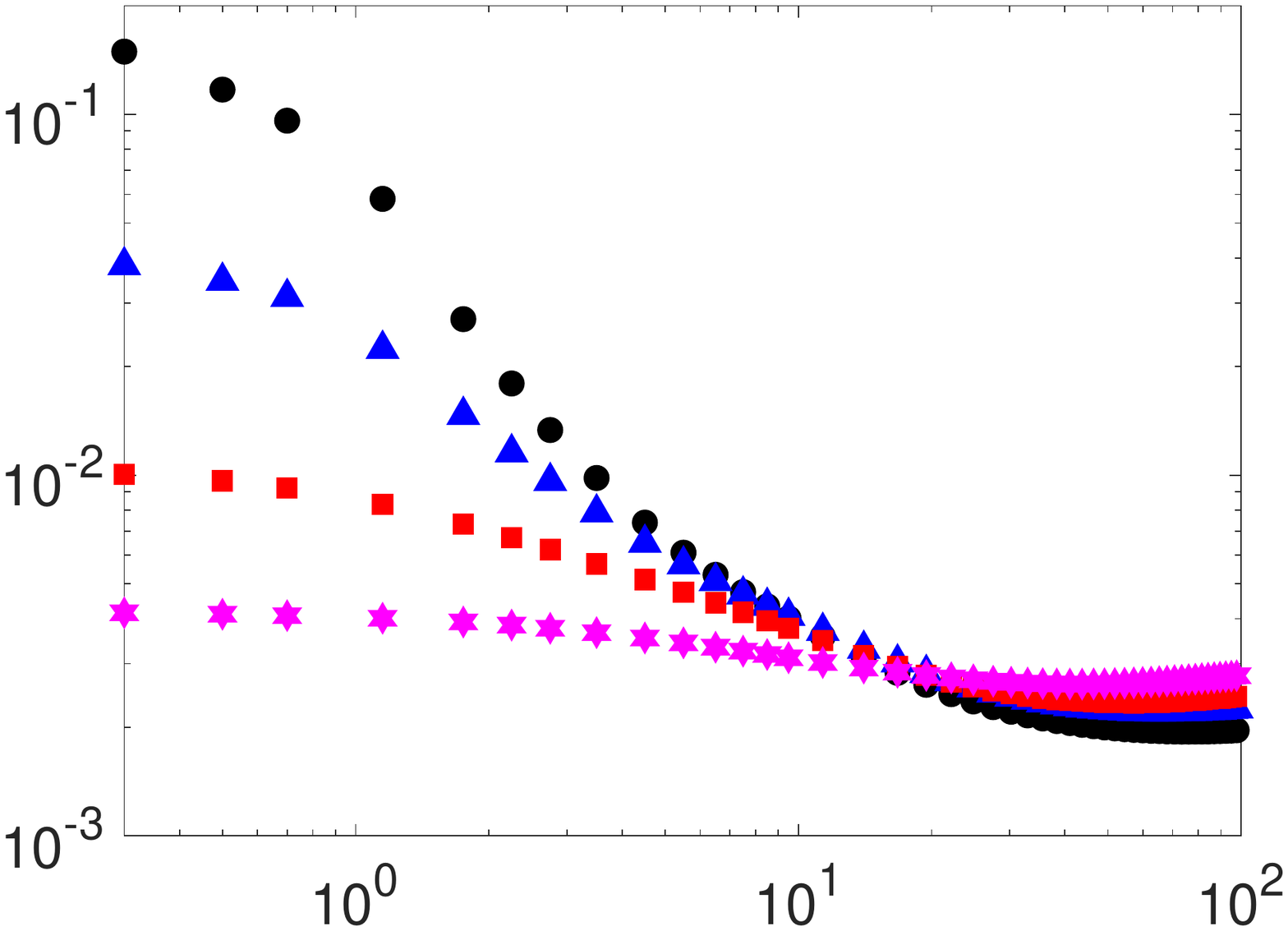}
\put(90,0){$z$}
\put(0,60){\rotatebox{90}{$\varrho$}}
\end{overpic}
\subfloat[]{}
\begin{overpic}
[trim = 0mm 60mm 0mm 50mm,scale=0.3,clip,tics=20]{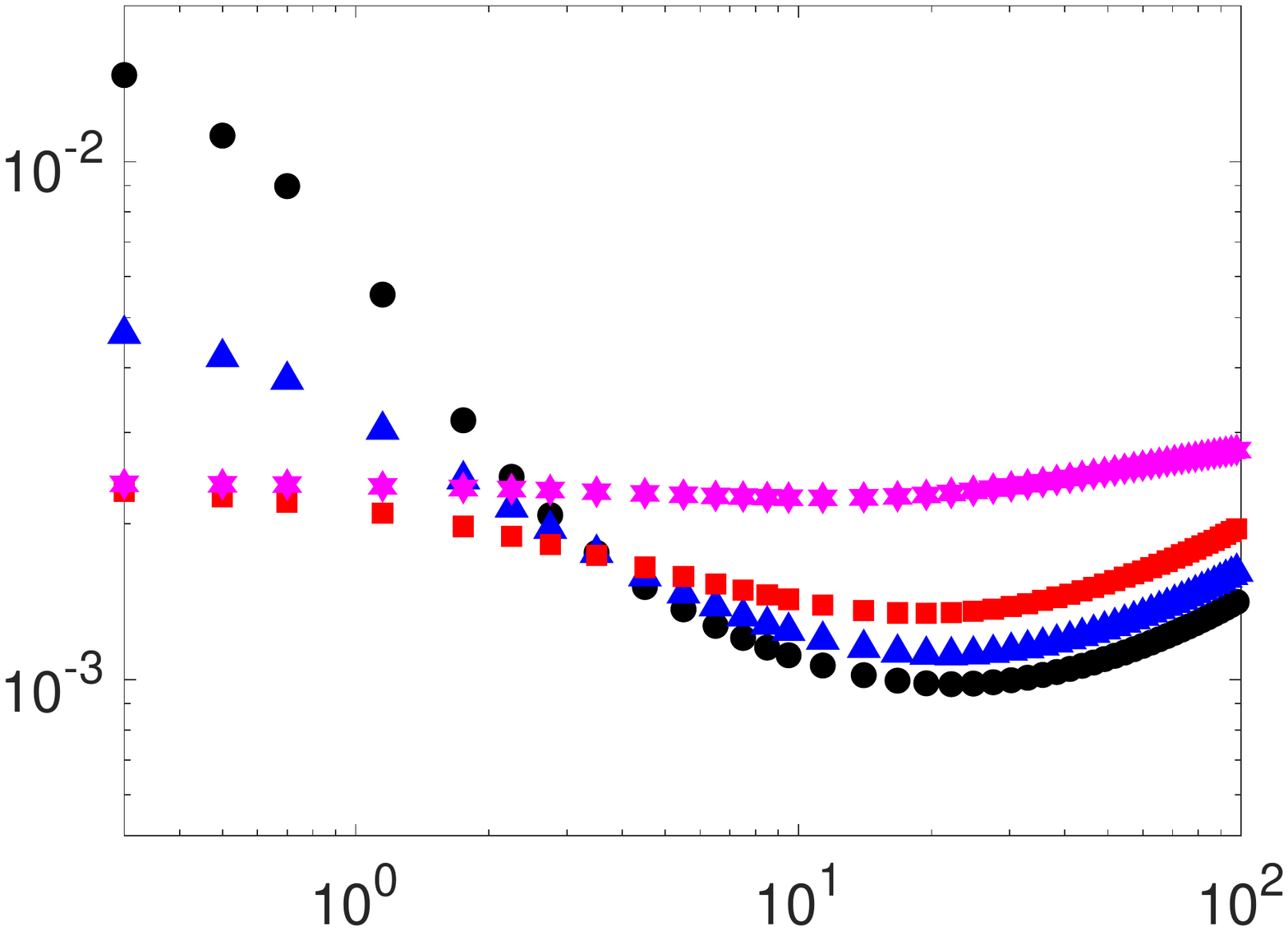}
\put(90,0){$z$}
\put(0,60){\rotatebox{90}{$\varrho$}}
\end{overpic}
\caption{Plot of $\varrho$ as a function of $z$ for different $Sv$, and (a) $St=0.93$, (b) $St=2.80$, (c) $St=9.30$, (d) $St=46.5$.}
\label{rho_plot}
\end{figure}
\FloatBarrier

We begin by considering in figure \ref{rho_plot} the results for concentration $\varrho(z)$ for different $Sv$ and $St$. For each $St$, the results show a very strong effect of $Sv$ on $\varrho$, with strong reductions in the near-wall concentration as $Sv$ is increased. Indeed, for the cases with $St=2.8, 9.3, 46.5$, the near wall value of $\varrho$ reduces by an order of magnitude as $Sv$ is increased from $Sv= 3 \times 10^{-4}$ to $Sv= 3 \times 10^{-2}$. This remarkable sensitivity of $\varrho$ to $Sv$ even when $Sv\ll1$ confirms the prediction of the analysis in \S\ref{TandA}. Conceptually, that gravity leads to a reduction in the value of the concentration near the wall is because with an absorbing wall, settling reduces the residence time of the particles in the near-wall region as their vertical velocity through the viscous sublayer is larger than it would be in the absence of settling. For the case of particles which elastically collide with the wall, the steady-state would be characterized by zero flux ($\langle w^p(t)\rangle_z=0$), and settling would actually lead to an increase of the near wall concentration in general since their weight would keep them trapped in the near wall region until they experience large enough fluctuations of the vertical fluid velocity to resuspend them into the flow. The analysis of \S\ref{TandA} can also be extended to this zero-flux case and would again show that even if $Sv\ll1$, settling can make a non-negligible contribution to the near wall concentration of the particles (this has been confirmed in results not shown here).

Not only do the results in figure \ref{rho_plot} show a strong effect of settling on the magnitude of the particle concentrations even when $Sv\ll1$, but they also indicate that settling affects the scaling behavior of $\varrho(z)$. In particular, as discussed in \S\ref{TandA}, the asymptotic analysis of \citet{sikovsky14} shows that for $Sv=0$, as the wall is approached the concentration behaves as a power law 
$\varrho(z)\sim z^{-\gamma}$, which was also confirmed using DNS data. Unlike those DNS results (or those in \citet{johnson20}), the results in figure \ref{rho_plot} show that as $z$ decreases, there is a noticeable kink in the curves for $\varrho$ around $z=O(1)$, below which $\varrho$ appears to flatten out. This change in behavior is due to the role of $Sv$ in \eqref{weq}.

\begin{figure}
\centering
\subfloat[]{}
\begin{overpic}
[trim = 0mm 60mm 0mm 50mm,scale=0.3,clip,tics=20]{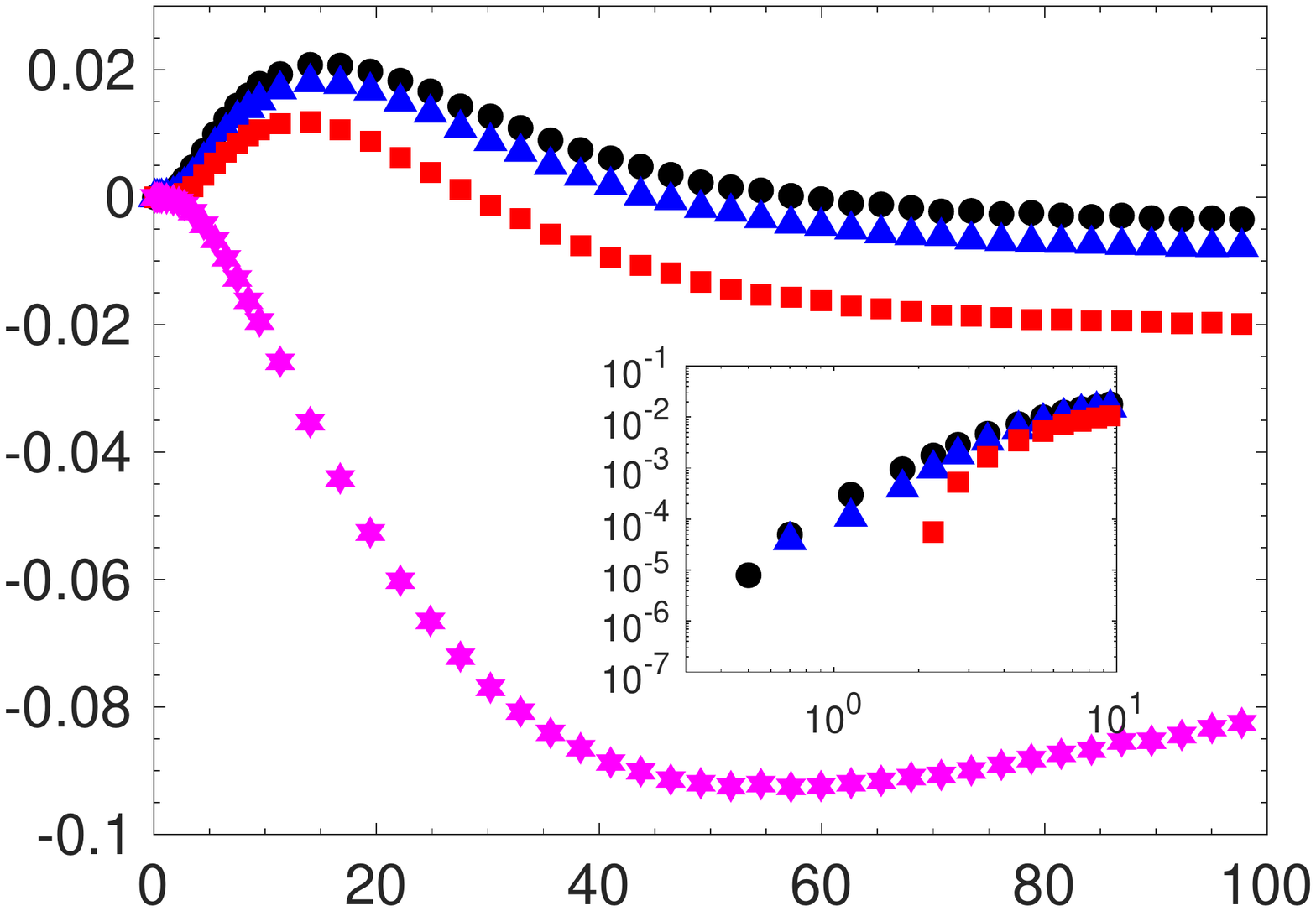}
\put(90,-2){$z$}
\put(-10,50){\rotatebox{90}{$\langle{u}^p(t)\rangle_{{z}}$}}
\end{overpic}
\subfloat[]{}
\begin{overpic}
[trim = 0mm 60mm 0mm 50mm,scale=0.3,clip,tics=20]{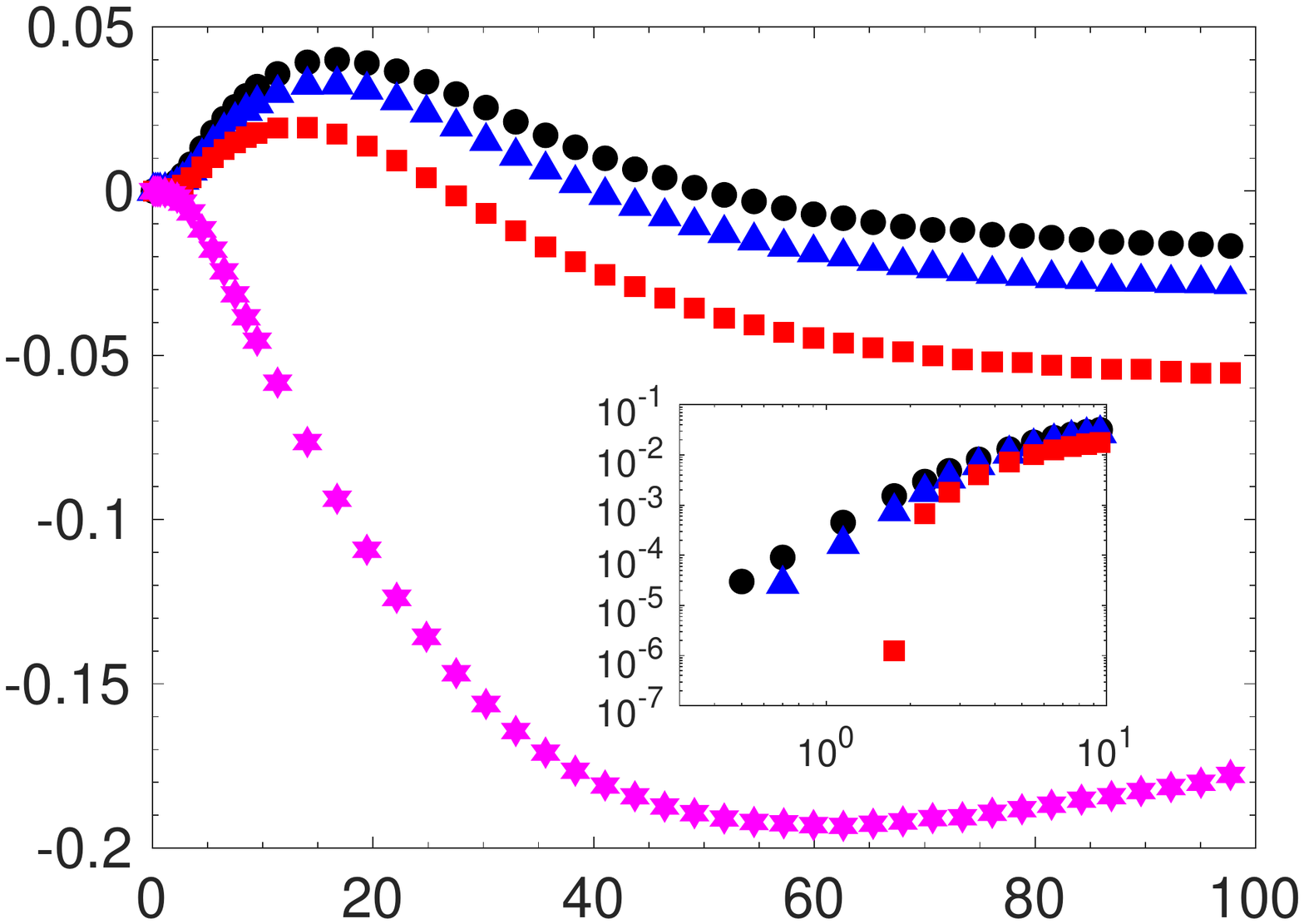}
\put(90,-2){$z$}
\put(-10,50){\rotatebox{90}{$\langle{u}^p(t)\rangle_{{z}}$}}
\end{overpic}\\\vspace{-4mm}
\subfloat[]{}
\begin{overpic}
[trim = 0mm 60mm 0mm 50mm,scale=0.3,clip,tics=20]{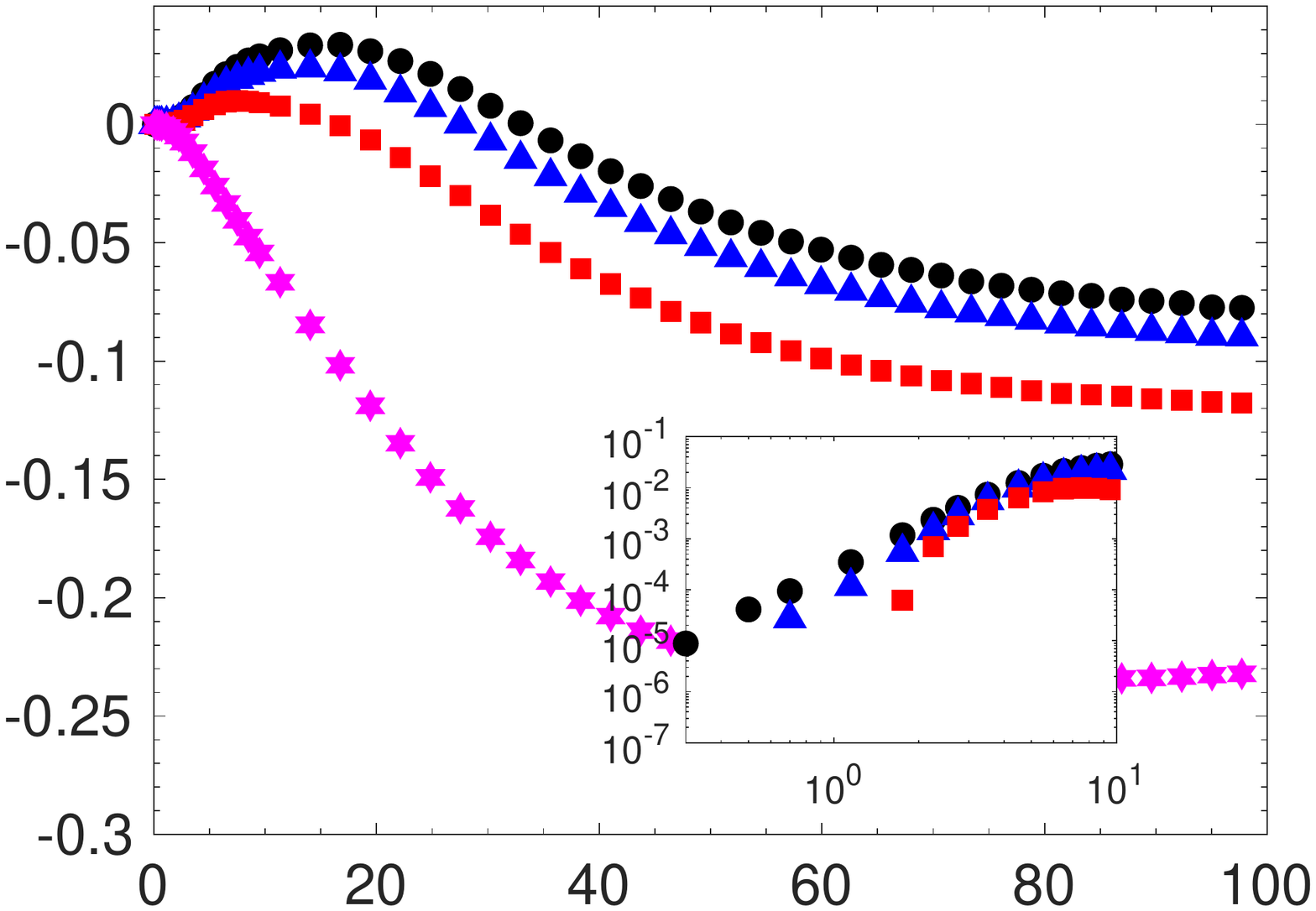}
\put(90,-2){$z$}
\put(-10,50){\rotatebox{90}{$\langle{u}^p(t)\rangle_{{z}}$}}
\end{overpic}
\subfloat[]{}
\begin{overpic}
[trim = 0mm 60mm 0mm 50mm,scale=0.3,clip,tics=20]{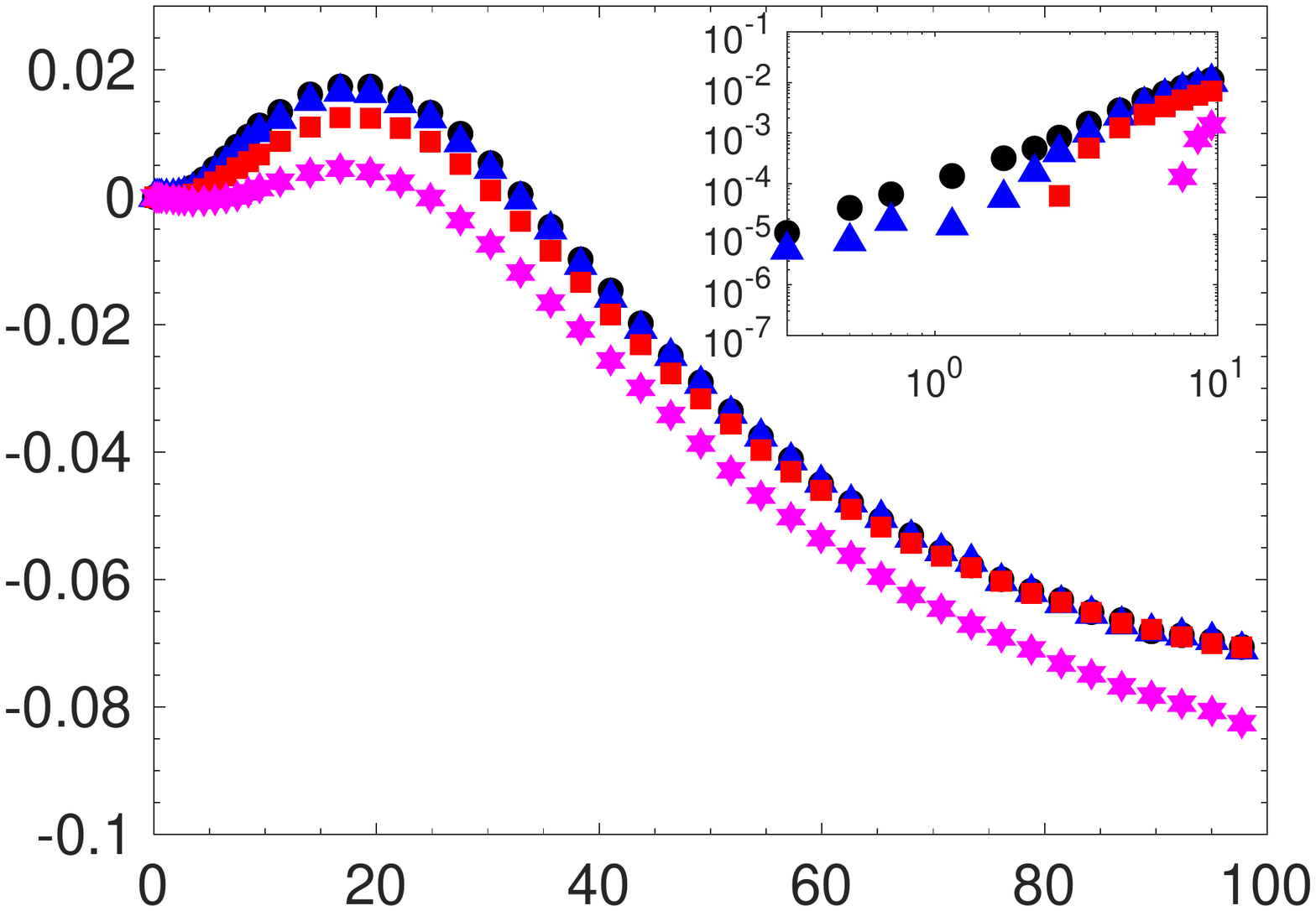}
\put(90,-2){$z$}
\put(-10,50){\rotatebox{90}{$\langle{u}^p(t)\rangle_{{z}}$}}
\end{overpic}
\caption{Plots of the average fluid velocity sampled by the particles $\langle{u}^p(t)\rangle_{{z}}$ as a function of $z$ for different $Sv$, and (a) $St=0.93$, (b) $St=2.80$, (c) $St=9.30$, (d) $St=46.5$. The insets correspond to the same plots but in a log-log scale to highlight the behavior for small $z$ (and in these plots the symbols disappear for small $z$ once the quantity becomes negative). Legend is the same as that in figure~\ref{rho_plot}.}
\label{PS_plot}
\end{figure}
\FloatBarrier
We now turn to consider how settling affects each of the terms in $\Lambda(z)$ as defined in \eqref{Lambda_eq}. In figure~\ref{PS_plot}, the results for $\langle{u}^p(t)\rangle_{{z}}$ are shown for each of the $St, Sv$ combinations. In the absence of settling, this term is known to be positive in the near wall region \cite{johnson20}, which is associated with the preference of the particles to accumulate in the near-wall ejection events of the turbulent boundary layer \cite{rashidi90,eaton94,marchioli02}. Our results also show $\langle{u}^p(t)\rangle_{{z}}>0$ near the wall for the smallest $Sv$ case $Sv=3 \times 10^{-4}$. However, as $Sv$ is increased, the size of $\langle{u}^p(t)\rangle_{{z}}$ reduces, and is negative near the wall for $Sv=3 \times 10^{-1}$. Moreover, the inset plots in figure~\ref{PS_plot} indicate that even for $Sv=3 \times 10^{-2}$, $\langle{u}^p(t)\rangle_{{z}}$ is negative for $z\leq O(1)$ (at which point the curve disappears on the logarithmic plot). Hence the preferential sampling of ejection events in the boundary layer by the inertial particles is very sensitive to $Sv$ even when $Sv\ll1$, and transitions to a preferential sampling of sweep events as $Sv$ increases (over a range of $z$ that depends on $St$ and $Sv$). This striking change in behavior can be understood as being due to the onset of the preferential sweeping mechanism \cite{maxey87,tom19} as $Sv$ increases, and has been seen in other DNS of particles settling through wall-bounded turbulence \cite{LeeLeeJFM2019}. This change in the sign of $\langle{u}^p(t)\rangle_{{z}}$ as $Sv$ increases also means that the role of this term in governing the particle concentration changes. When $\langle{u}^p(t)\rangle_{{z}}>0$ this term hinders the settling of the particles towards the wall and acts to reduce their near wall accumulation \cite{johnson20}. However, when $\langle{u}^p(t)\rangle_{{z}}<0$, this term contributes to the settling of the particles towards the wall and so can contribute to their near wall accumulation. 
\begin{figure}
\centering
\subfloat[]{}
\begin{overpic}
[trim = 0mm 60mm 0mm 50mm,scale=0.3,clip,tics=20]{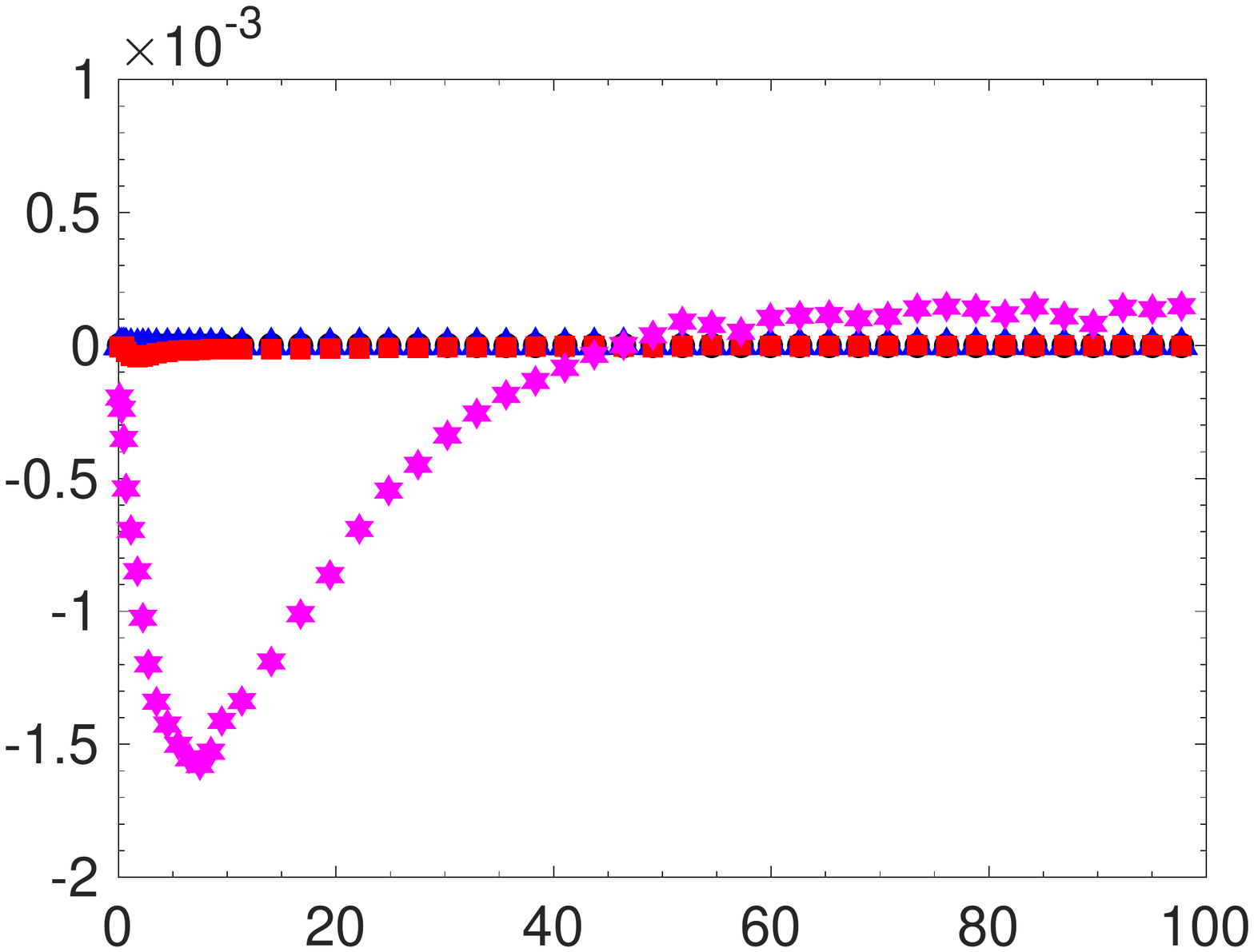}
\put(90,-2){$z$}
\put(-10,50){\rotatebox{90}{$-St D_t\langle w^p(t)\rangle_z$}}
\end{overpic}
\subfloat[]{}
\begin{overpic}
[trim = 0mm 60mm 0mm 50mm,scale=0.3,clip,tics=20]{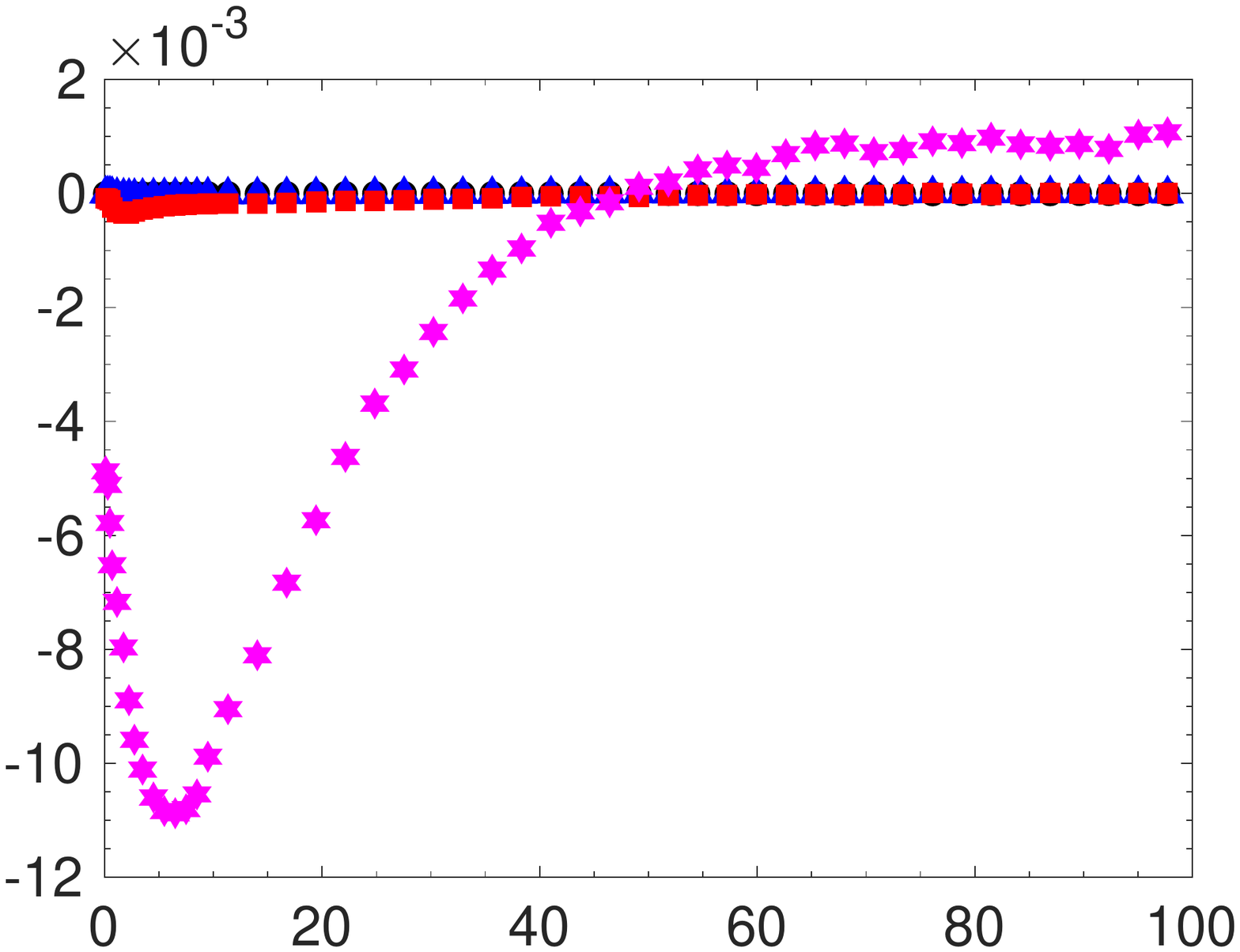}
\put(90,-2){$z$}
\put(-10,50){\rotatebox{90}{$-St D_t\langle w^p(t)\rangle_z$}}
\end{overpic}\\\vspace{-4mm}
\subfloat[]{}
\begin{overpic}
[trim = 0mm 60mm 0mm 50mm,scale=0.3,clip,tics=20]{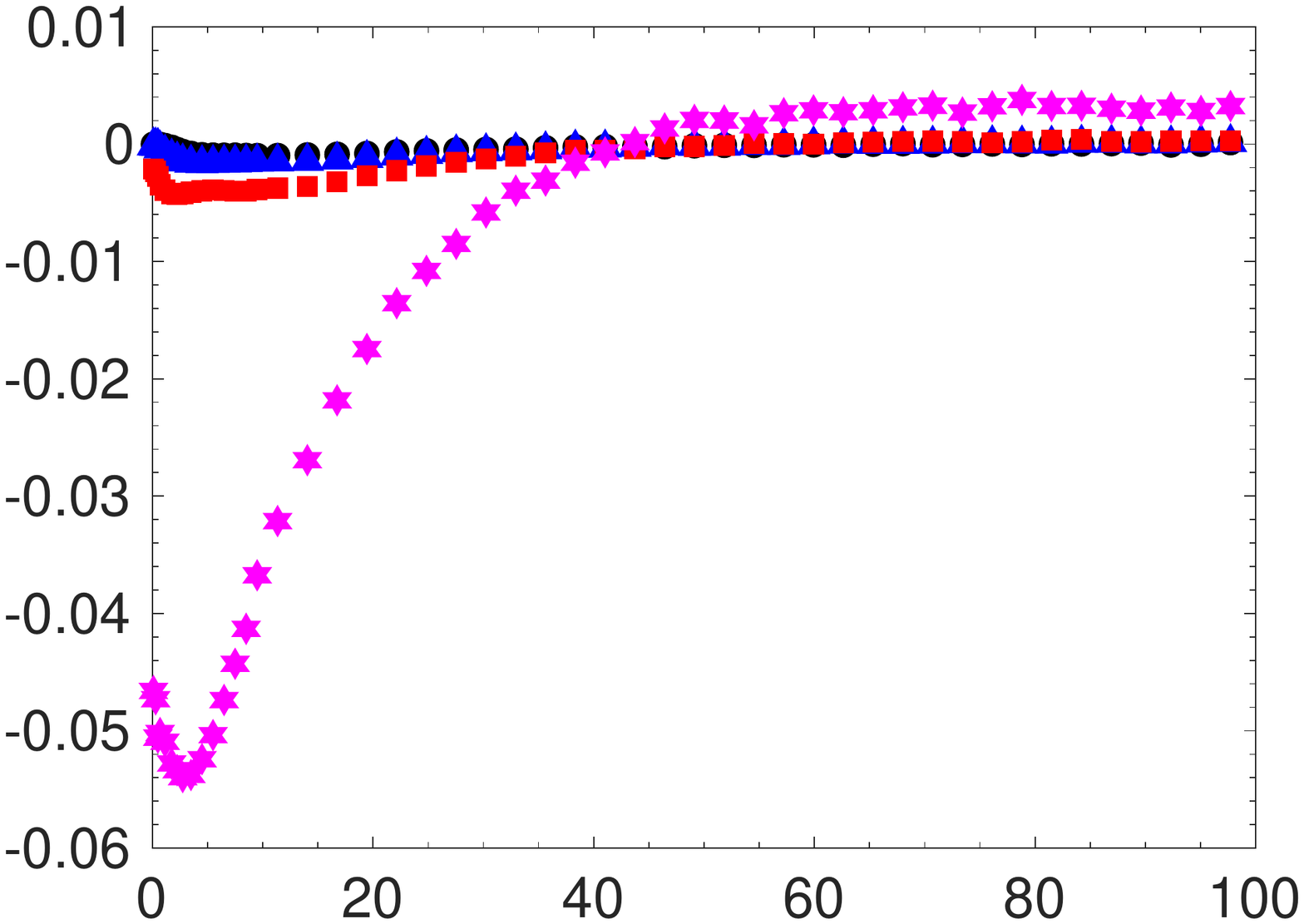}
\put(90,-2){$z$}
\put(-10,50){\rotatebox{90}{$-St D_t\langle w^p(t)\rangle_z$}}
\end{overpic}
\subfloat[]{}
\begin{overpic}
[trim = 0mm 60mm 0mm 50mm,scale=0.3,clip,tics=20]{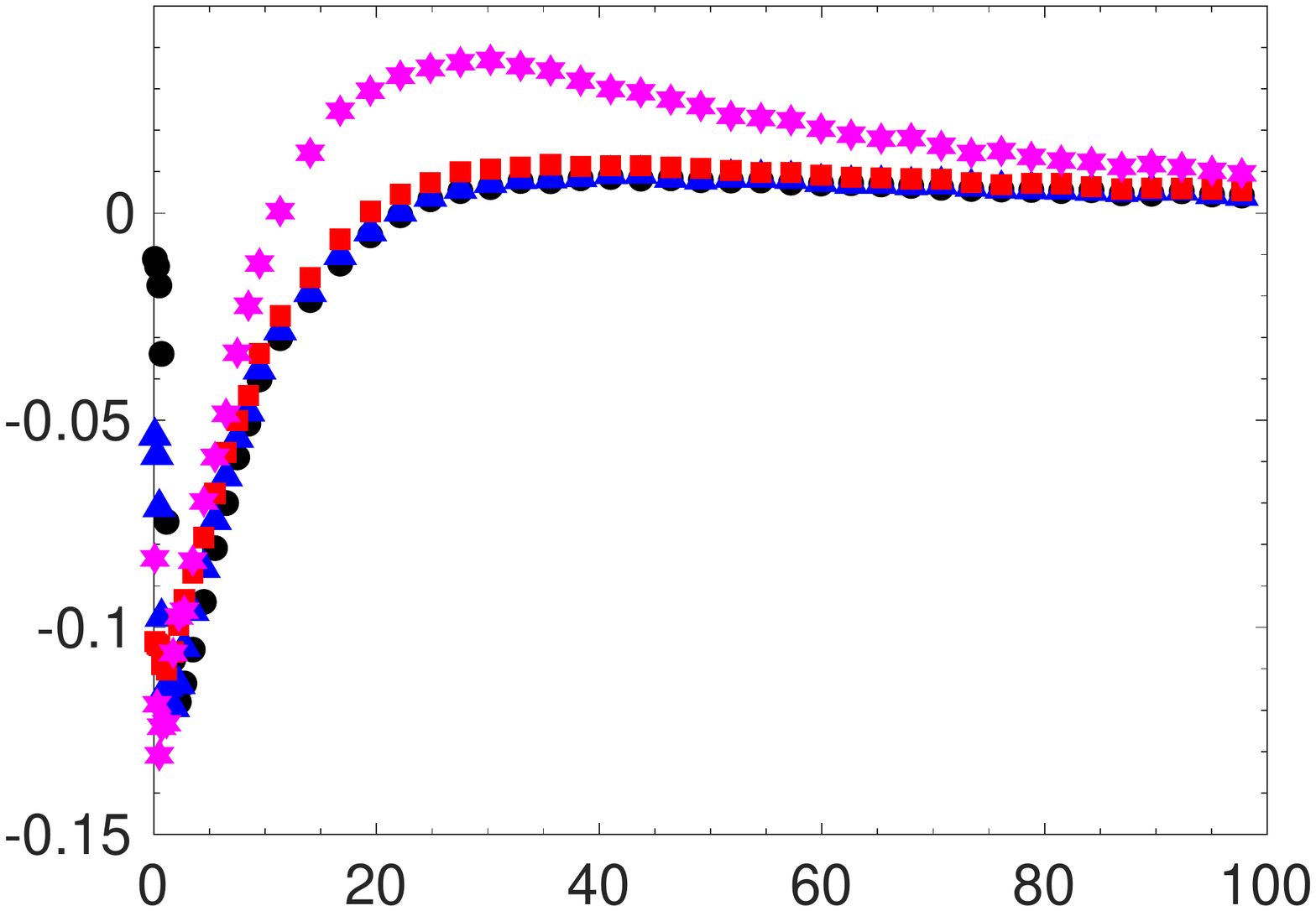}
\put(90,-2){$z$}
\put(-10,50){\rotatebox{90}{$-St D_t\langle w^p(t)\rangle_z$}}
\end{overpic}
\caption{Plots of the average vertical particle velocity $\langle{w}^p(t)\rangle_{{z}}$ as a function of $z$ for different $Sv$, and (a) $St=0.93$, (b) $St=2.80$, (c) $St=9.30$, (d) $St=46.5$. Legend is the same as that in figure~\ref{rho_plot}. Note that inset plots showing the results in a log-log scale are not shown for this quantity as the data is very noisy for small $z$ due to the spatial differentiation of the data for $\langle w^p(t)\rangle_z$ involved in computing $D_t\langle w^p(t)\rangle_z$.}
\label{Acc_plot}
\end{figure}
\FloatBarrier
In figure~\ref{Acc_plot}, the results for $-St D_t\langle w^p(t)\rangle_z$ are shown for each of the $St, Sv$ combinations. Since the flow has constant negative flux $\varrho\langle w^p(t)\rangle_z<0$, then since $\varrho$ increases as the wall is approached, it follows that $-St D_t\langle w^p(t)\rangle_z$ is negative, as shown in the results. In \citet{bragg2021mechanisms} it was shown that for $Sv=3\times 10^{-2}$ this acceleration term makes a negligible contribution to $\langle w^p(t)\rangle_z$. Our results for different $Sv$ reveal that although this term generally makes a small contribution to $\langle w^p(t)\rangle_z$, it can become important for large $St$ as $Sv$ increases. However, for these larger $St,Sv$ cases, the concentration is almost uniform (see figure \ref{rho_plot}), and so while $-St D_t\langle w^p(t)\rangle_z$ can be important for predicting the settling velocity $\langle w^p(t)\rangle_z$, it is not important in the regime of $St,Sv$ for which there is a significant build up of particle concentration near the wall.
\begin{figure}
\centering
\subfloat[]{}
\begin{overpic}
[trim = 0mm 60mm 0mm 50mm,scale=0.3,clip,tics=20]{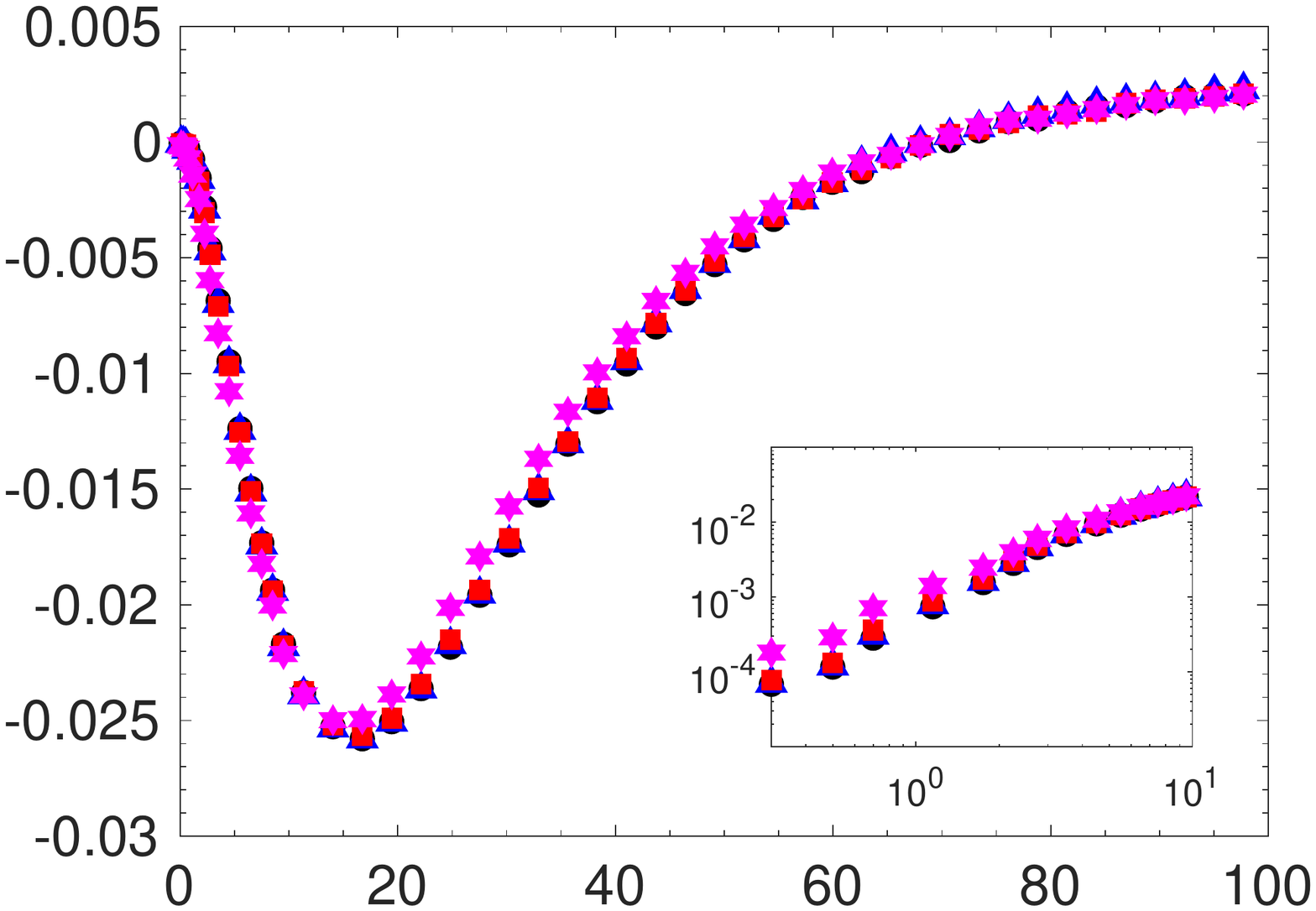}
\put(90,-2){$z$}
\put(-8,45){\rotatebox{90}{$-St\nabla_z \mathcal{W}$}}
\end{overpic}
\subfloat[]{}
\begin{overpic}
[trim = 0mm 60mm 0mm 50mm,scale=0.3,clip,tics=20]{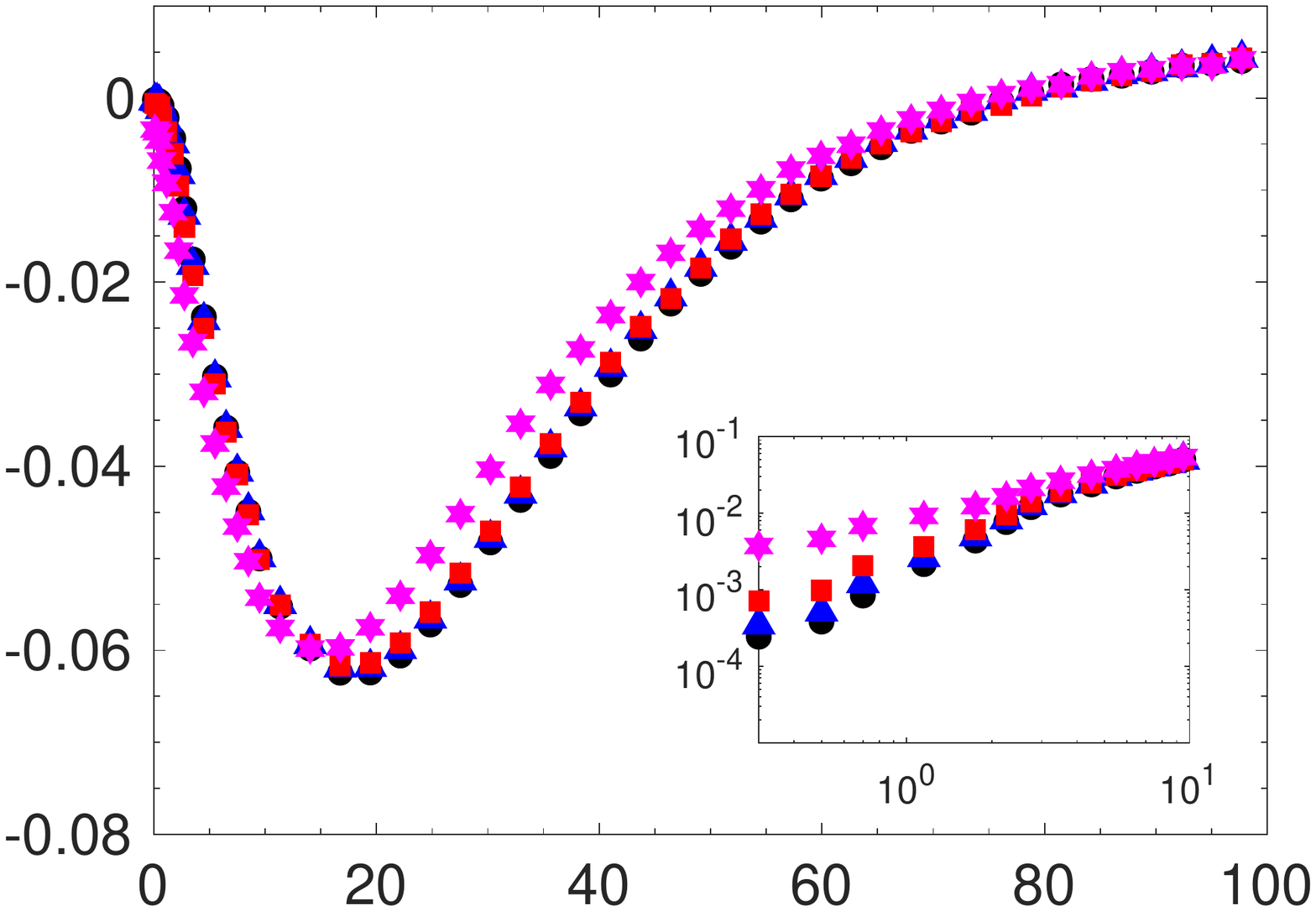}
\put(90,-2){$z$}
\put(-8,45){\rotatebox{90}{$-St\nabla_z \mathcal{W}$}}
\end{overpic}\\\vspace{-4mm}
\subfloat[]{}
\begin{overpic}
[trim = 0mm 60mm 0mm 50mm,scale=0.3,clip,tics=20]{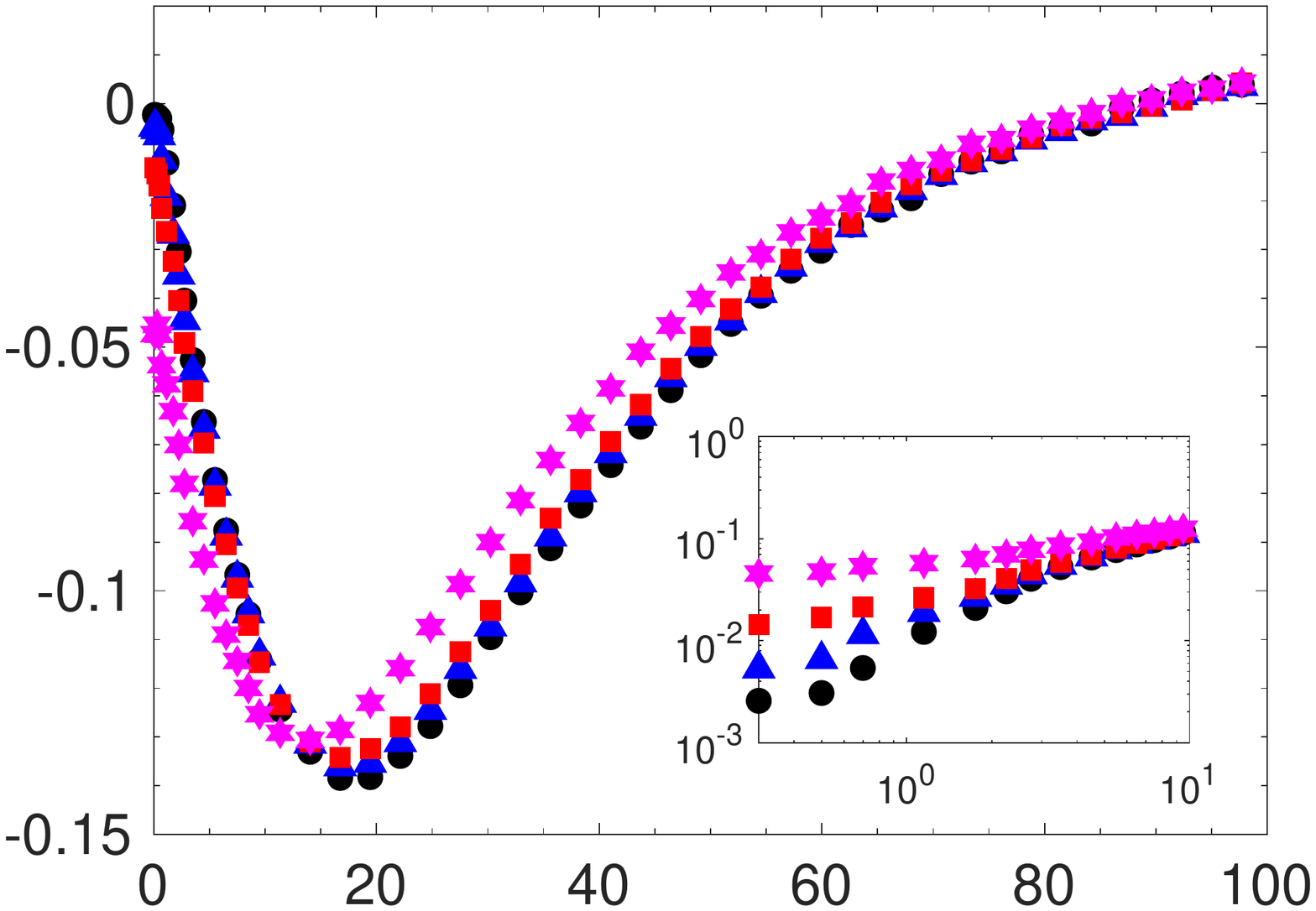}
\put(90,-2){$z$}
\put(-4,45){\rotatebox{90}{$-St\nabla_z \mathcal{W}$}}
\end{overpic}
\subfloat[]{}
\begin{overpic}
[trim = 0mm 60mm 0mm 50mm,scale=0.3,clip,tics=20]{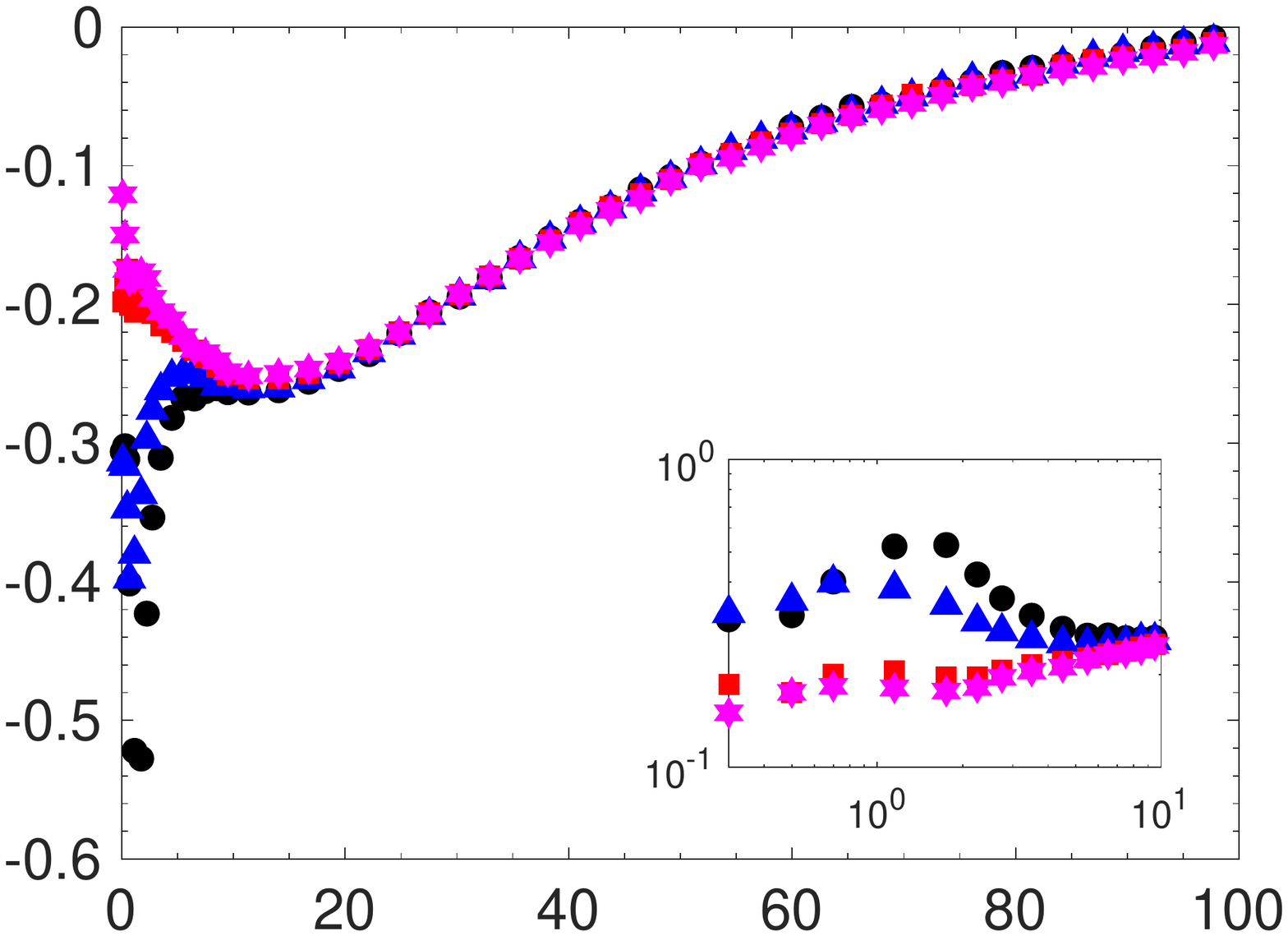}
\put(90,-2){$z$}
\put(-4,45){\rotatebox{90}{$-St\nabla_z \mathcal{W}$}}
\end{overpic}
\caption{Plots of the turbophoretic velocity $-St\nabla_z \mathcal{W}$ as a function of $z$ for different $Sv$, and (a) $St=0.93$, (b) $St=2.80$, (c) $St=9.30$, (d) $St=46.5$. The insets correspond to the same plots (except that the vertical axis is now $St\nabla_z \mathcal{W}$) but in a log-log scale to highlight the behavior for small $z$. Legend is the same as that in figure~\ref{rho_plot}.}
\label{TD_plot}
\end{figure}
\FloatBarrier
In figure \ref{TD_plot}, the results for the turbophoretic drift velocity $-St\nabla_z \mathcal{W}$ are shown for different $St$ and $Sv$. Compared to the results for $\langle{u}^p(t)\rangle_{{z}}$ and $-St D_t\langle w^p(t)\rangle_z$, the results show that $-St\nabla_z \mathcal{W}$ is relatively insensitive to $Sv$. The insets to the plots reveal, however, that there is a strong dependency on $Sv$ in the near wall region $z\leq O(1)$, with $-St\nabla_z \mathcal{W}$ increasing as $Sv$ increases for $Sv< 3\times 10^{-1}$. The corresponding results for $\mathcal{W}$ (not shown) also show that $\mathcal{W}$ increases as $Sv$ increases. This behavior may be partially understood by considering the expression for $\mathcal{W}$ obtained using the formal solution to \eqref{eom}, namely
\begin{align}
    \mathcal{W}(z)=\int^0_{-\infty}\int^0_{-\infty}e^{(s+s')/St}\Big(\langle u^p(s)u^p(s')\rangle_z-\langle u^p(s)\rangle_z\langle u^p(s')\rangle_z\Big)\,ds'\,ds.\label{Wsol}
\end{align}
Settling reduces the correlation timescale of the flow seen by the particle \cite{csanady}, meaning that increasing $Sv$ will reduce the time span $|s-s'|$ over which the covariance $\langle u^p(s)u^p(s')\rangle_z$ is finite. This in turn would have the effect of reducing $\mathcal{W}$ with increasing $Sv$ (see \cite{ireland16b} for the discussion of an analogous effect in the context of how settling impacts particle-pair relative velocities). The results in figure~\ref{PS_plot} show, however, that increasing $Sv$ leads to a reduction in the preferential sampling of the flow for $Sv\leq 3\times 10^{-2}$. If this reduction in the preferential sampling leads to a greater reduction in $\langle u^p(s)\rangle_z\langle u^p(s')\rangle_z$ than the reduction in the correlation timescale associated with $\langle u^p(s)u^p(s')\rangle_z$, then according to \eqref{Wsol} the overall effect of increasing $Sv$ would be to increase $\mathcal{W}$, as observed. Such an argument also shows the way in which the preferential sampling of the flow impacts $\langle w^p(t)\rangle_z$ not only explicitly through the term $\langle u^p(t)\rangle_z$ in \eqref{weq}, but also implicitly through its effect on $-St\nabla_z \mathcal{W}$ which itself depends on how the particles interact with and sample the flow.

\begin{figure}
\centering
\subfloat[]{}
\begin{overpic}
[trim = 0mm 60mm 0mm 50mm,scale=0.3,clip,tics=20]{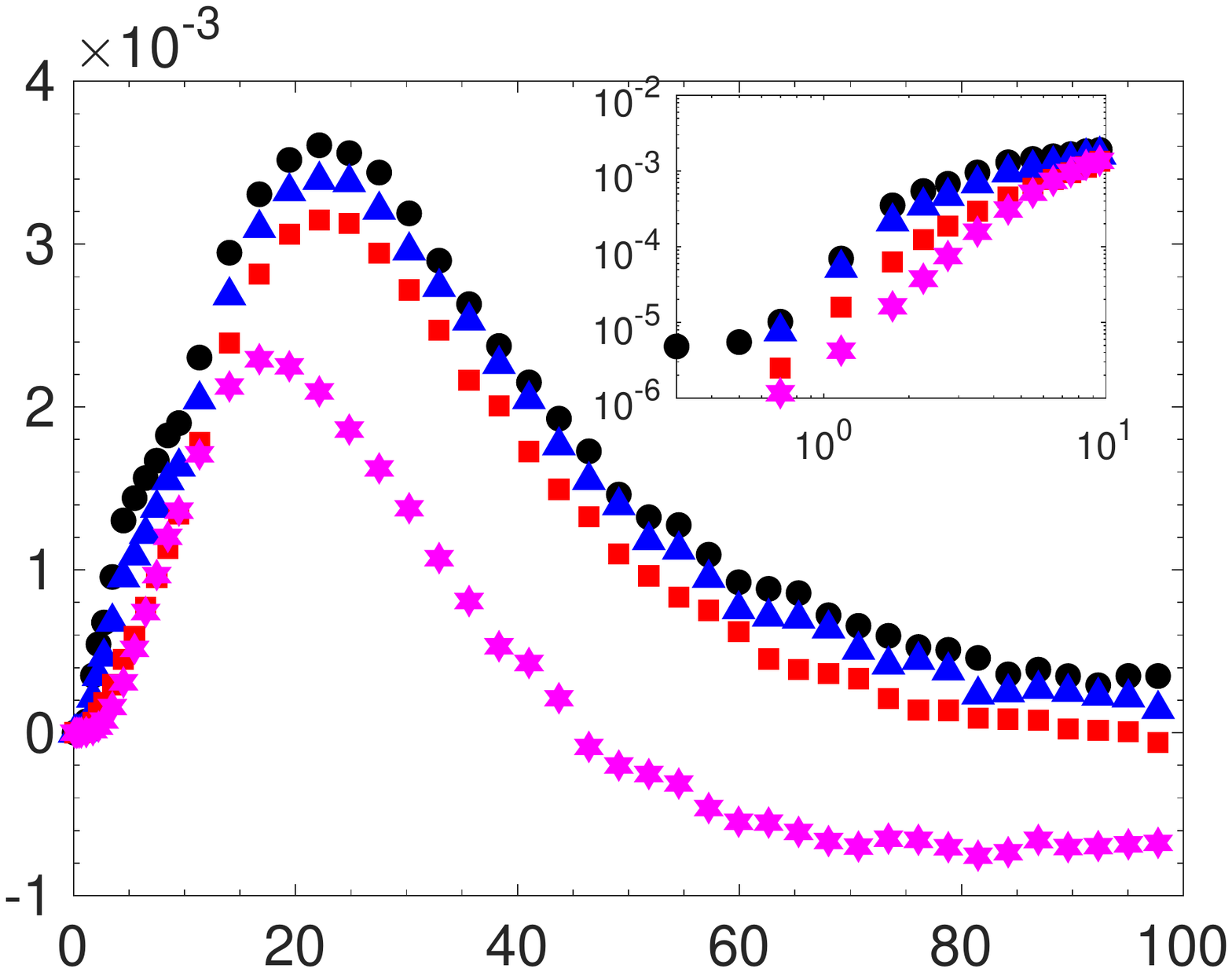}
\put(90,-2){$z$}
\put(-4,45){\rotatebox{90}{$-St\mathcal{W}\varrho^{-1}\nabla_z \varrho$}}
\end{overpic}
\subfloat[]{}
\begin{overpic}
[trim = 0mm 60mm 0mm 50mm,scale=0.3,clip,tics=20]{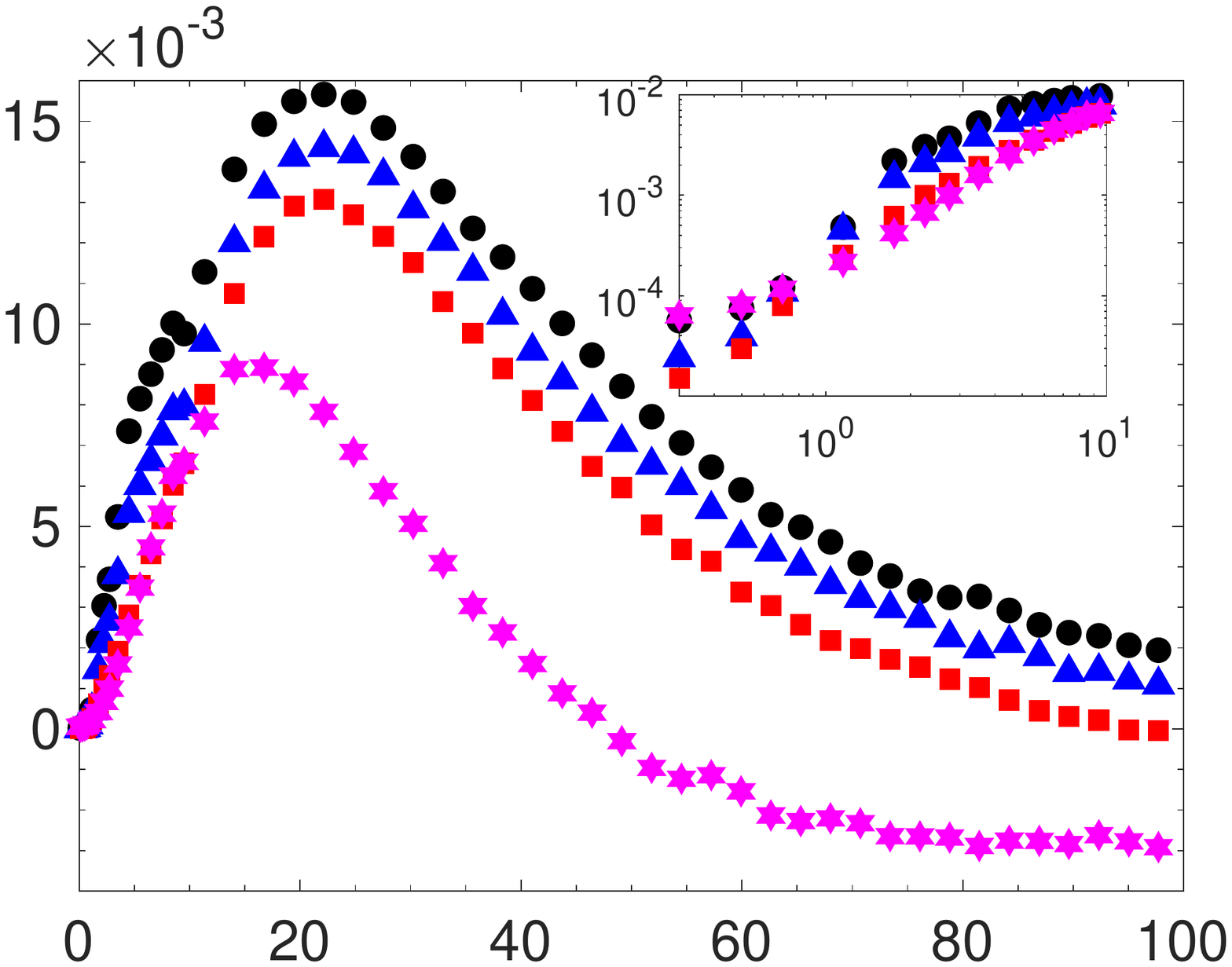}
\put(90,-2){$z$}
\put(-4,45){\rotatebox{90}{$-St\mathcal{W}\varrho^{-1}\nabla_z \varrho$}}
\end{overpic}\\\vspace{-4mm}
\subfloat[]{}
\begin{overpic}
[trim = 0mm 60mm 0mm 50mm,scale=0.3,clip,tics=20]{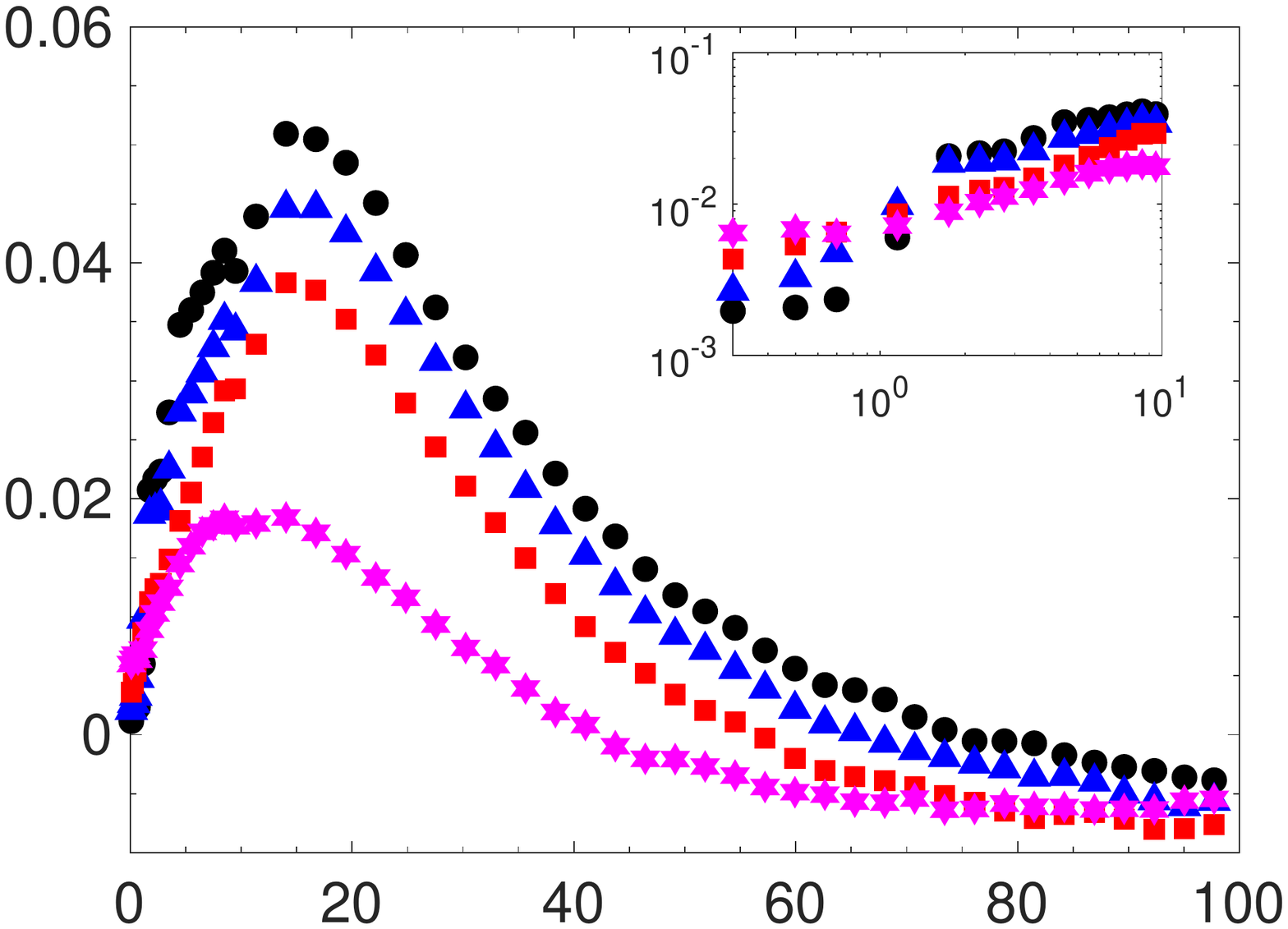}
\put(90,-2){$z$}
\put(-4,45){\rotatebox{90}{$-St\mathcal{W}\varrho^{-1}\nabla_z \varrho$}}
\end{overpic}
\subfloat[]{}
\begin{overpic}
[trim = 0mm 60mm 0mm 50mm,scale=0.3,clip,tics=20]{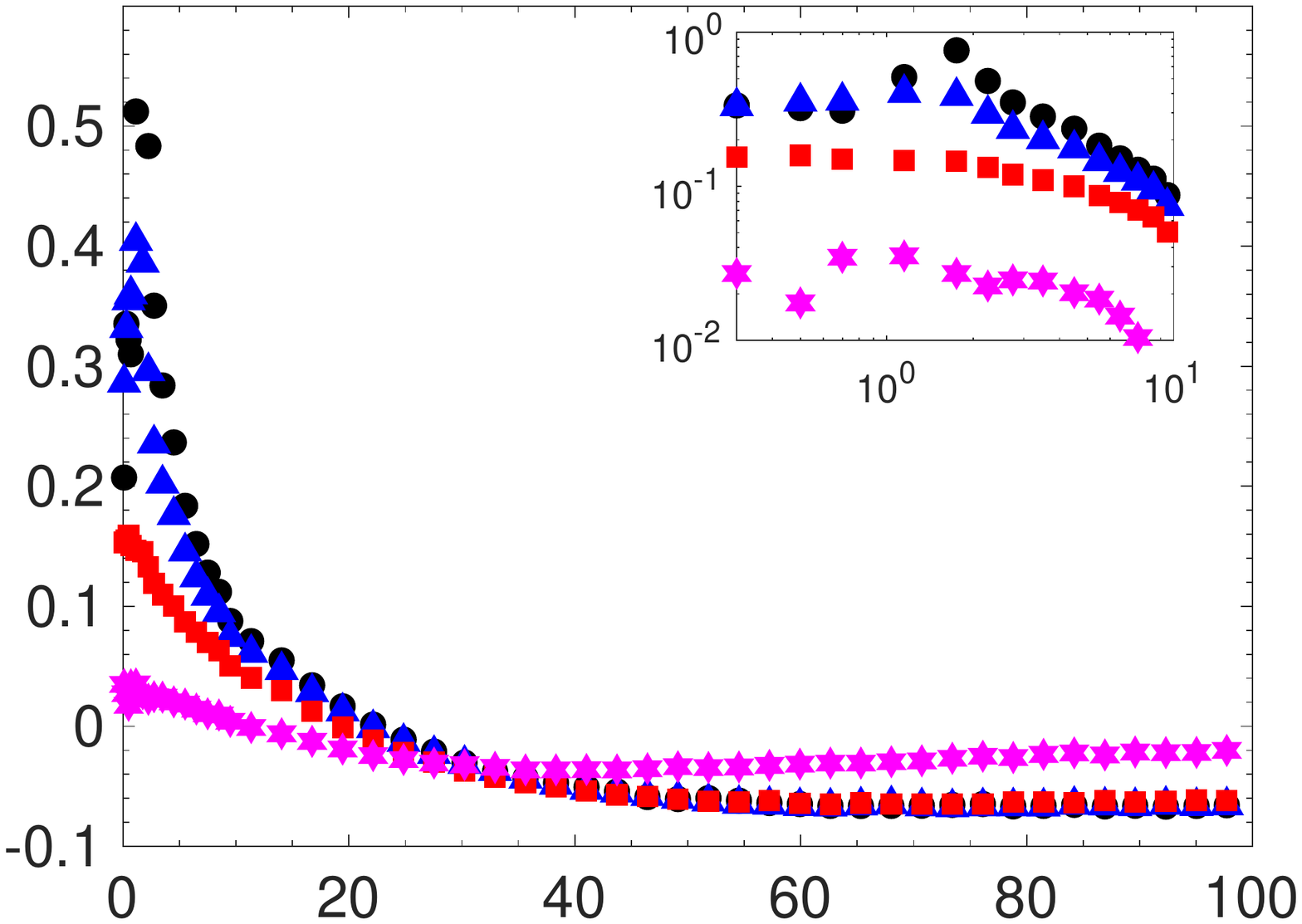}
\put(90,-2){$z$}
\put(-4,45){\rotatebox{90}{$-St\mathcal{W}\varrho^{-1}\nabla_z \varrho$}}
\end{overpic}
\caption{Plots of the diffusive velocity $-St\mathcal{W}\varrho^{-1}\nabla_z \varrho$ as a function of $z$ for different $Sv$, and (a) $St=0.93$, (b) $St=2.80$, (c) $St=9.30$, (d) $St=46.5$. The insets correspond to the same plots but in a log-log scale to highlight the behavior for small $z$. Legend is the same as that in figure~\ref{rho_plot}.}
\label{Diff_plot}
\end{figure}

Finally, in figure \ref{Diff_plot}, the results for the diffusive  velocity $-St\mathcal{W}\varrho^{-1}\nabla_z \varrho$ are shown for different $St$ and $Sv$. Near the wall this velocity is positive since $\nabla_z\varrho<0$ there, so that this diffusive velocity hinders the settling velocity of the particles. This term is also very sensitive to $Sv$ even for $Sv\ll1$, which is mainly due to the sensitivty of $\varrho$ to $Sv$ that was shown in figure \ref{rho_plot}. Throughout much of the boundary layer this term makes a sub-leading contribution in \eqref{weq}. However, comparing the logarithmic scale inset plots to those of the other quantities reveals that very close to the wall where the gradients of $\varrho$ are strongest, this term can be significant compared to the other contributions in \eqref{weq}.
\FloatBarrier

\section{Conclusions}
We have considered the role of gravitational settling on the concentration profiles of small inertial particles in  wall-bounded turbulent flows. We provided theoretical arguments and DNS results that show that settling can play a leading order role in determining the concentrations, even when the Stokes settling velocity is very small compared with the fluid friction velocity. The reason is that the dynamical relevance of settling is determined by the size of the Stokes settling velocity compared with the other mechanisms contributing to the particle vertical velocity, not compared with the fluid friction velocity (or any other fluid velocity scale). In the theoretical analysis, this corresponds to saying that settling can only be ignored if $\Lambda_0\equiv \Lambda\vert_{Sv=0}$ (see \eqref{Lambda_eq} for the definition of $\Lambda(z)$) is much larger than $O(Sv)$. However, as we have shown, this condition will always be violated for $Sv>0$ in some region close to the wall since in the viscous sub-layer, $\lim_{z\to0}\Lambda_0\to 0$. 

Quantitatively, the DNS results showed that the near-wall concentration is highly dependent on $Sv$ even when $Sv\ll 1$, and indeed the concentration can be reduced by an order of magnitude when $Sv$ is increased from $O(10^{-4})$ and $O(10^{-2})$. The results also show that the preferential sampling of ejection events in the boundary layer by inertial particles that occurs for $Sv=0$ is profoundly altered as $Sv$ is increased, and is replaced by a preferential sampling of sweeping events due to the onset of the preferential sweeping mechanism. The results are very consequential for understanding and predicting the concentration profiles of inertial particles in wall-bounded turbulent flows, since many previous studies neglected the effect of settling under the assumption that its effect is negligible for $Sv\ll1$.

One practical issue is that it is not always possible to reliably predict $\Lambda_0(z)$ \textit{a priori}, and therefore when performing numerical simulations, it may not be clear whether the condition $\Lambda_0\gg O(Sv)$ will be satisfied so as to justify neglecting particle settling. Our results suggest that it is probably best to always retain the particle settling, and then one can check \textit{a posteriori} whether the settling plays any important role.

\section*{Acknowledgements}

The authors acknowledge grant G00003613-ArmyW911NF-17-0366 from the US Army Research Office. Computational resources were provided by the High Performance Computing Modernization Program (HPCMP), and by the Center for Research Computing (CRC) at the University of Notre Dame.

The authors report no conflict of interest.

%


\bibliography{PRF_version}

\end{document}